\documentclass[
reprint,
superscriptaddress,
showpacs,
preprintnumbers,
nofootinbib,
amsmath,
amssymb,
aps,
prd,
floatfix
]{revtex4-2}

\usepackage[utf8]{inputenc}
\usepackage[normalem]{ulem}
\usepackage{graphicx}
\usepackage{dcolumn}
\usepackage{bm}
\usepackage{color}
\usepackage{xcolor}
\usepackage[colorlinks=true,allcolors=purple]{hyperref}
\usepackage{url}
\usepackage{enumerate}

\usepackage{slashed}
\usepackage{multirow}
\usepackage{mathrsfs} 
\usepackage{amsmath}

\usepackage{bbold}
\usepackage{mathrsfs}
\usepackage{braket}
\usepackage{physics}
\usepackage{multirow}

\usepackage{cleveref}

\usepackage{fontawesome} 
\definecolor{blue-violet}{rgb}{0.33, 0.17, 0.89}

\newcommand{\refeq}[1]{Eq.~(\ref{#1})}

\newcommand{\reffig}[1]{Fig.~\ref{#1}}

\newcommand{\refref}[1]{Ref.~\cite{#1}}
\newcommand{\refrefs}[2]{Refs.~\cite{#1}~and~\cite{#2}}

\def\muboone{MicroBooNE}

\newcounter{CommentCount}
\definecolor{MH}{rgb}{0.0,0.6,9}
\definecolor{palatinate}{rgb}{0.494, 0.192, 0.482}

\renewcommand{\phi}{\varphi}

\usepackage{siunitx}

\DeclareSIUnit \s {\second}
\DeclareSIUnit \ns {\nano\second}
\DeclareSIUnit \mus {\micro\second}
\DeclareSIUnit \ms {\milli\second}
\DeclareSIUnit \MB {\mega\byte}
\DeclareSIUnit \GB {\giga\byte}
\DeclareSIUnit \TB {\tera\byte}
\DeclareSIUnit \PB {\peta\byte}
\DeclareSIUnit \Mbps {\mega\bit/\s}
\DeclareSIUnit \Gbps {\giga\bit/\s}
\DeclareSIUnit \Tbps {\tera\bit/\s}
\DeclareSIUnit \Pbps {\peta\bit/\s}
\DeclareSIUnit \kton {\kilo\tonne} 
\DeclareSIUnit \kt {\kilo\tonne}
\DeclareSIUnit \Mt {\mega\tonne}
\DeclareSIUnit \eV {\electronvolt}
\DeclareSIUnit \keV {\kilo\electronvolt}
\DeclareSIUnit \MeV {\mega\electronvolt}
\DeclareSIUnit \GeV {\giga\electronvolt}
\DeclareSIUnit \TeV {\tera\electronvolt}
\DeclareSIUnit \PeV {\peta\electronvolt}
\DeclareSIUnit \EeV {\exa\electronvolt}
\DeclareSIUnit \m {\meter}
\DeclareSIUnit \cm {\centi\meter}
\DeclareSIUnit \in {\inchcommand}
\DeclareSIUnit \km {\kilo\meter}
\DeclareSIUnit \kV {\kilo\volt}
\DeclareSIUnit \kW {\kilo\watt}
\DeclareSIUnit \MW {\mega\watt}
\DeclareSIUnit \MHz {\mega\hertz}
\DeclareSIUnit \mrad {\milli\radian}
\DeclareSIUnit \year {years}
\DeclareSIUnit \POT {POT}
\DeclareSIUnit \sig {$\sigma$}
\DeclareSIUnit\parsec{pc}
\DeclareSIUnit\lightyear{ly}
\DeclareSIUnit\foot{ft}
\DeclareSIUnit\ft{ft}
\DeclareSIUnit \ppb{ppb}
\DeclareSIUnit \ppt{ppt}
\DeclareSIUnit \samples{S}
\DeclareSIUnit \pe{PE}
\DeclareSIUnit \T{T}

\newcommand{\enu}{\E_\enu}

\begin{document}

\preprint{\hfill FTPI-MINN-21-16}

\title{Heavy neutral leptons below the kaon mass at hodoscopic neutrino detectors}

\author{Carlos A. Arg{\"u}elles}
\email{carguelles@fas.harvard.edu}
\affiliation{Department of Physics \& Laboratory for Particle Physics and Cosmology, Harvard University, Cambridge, MA 02138, USA}

\author{Nicol\`o Foppiani}
\email{nicolofoppiani@g.harvard.edu}
\affiliation{Department of Physics \& Laboratory for Particle Physics and Cosmology, Harvard University, Cambridge, MA 02138, USA}

\author{Matheus Hostert}
\email{mhostert@perimeterinstitute.ca}
\affiliation{Perimeter Institute for Theoretical Physics, Waterloo, ON N2J 2W9, Canada}
\affiliation{School of Physics and Astronomy, University of Minnesota, Minneapolis, MN 55455, USA}
\affiliation{William I. Fine Theoretical Physics Institute, School of Physics and Astronomy, University of
Minnesota, Minneapolis, MN 55455, USA}

\date{\today}

\begin{abstract}
Heavy neutral leptons ($N$) below the kaon mass are severely constrained by cosmology and lab-based searches for their decays in flight.
If $N$ interacts via an additional force, $N\to\nu e^+e^-$ decays are enhanced and cosmological limits can be avoided.
We show that the T2K and MicroBooNE neutrino experiments provide the best limits on the mixing of $N$ with muon-neutrinos, outperforming past-generation experiments, previously thought to dominate.
We constrain models with electromagnetically-decaying and long-lived $N$, such as in a transition-magnetic-moment portal and in a leptophilic axion-like particle portal, invoked to explain the MiniBooNE excess.
By considering these models as representative examples, our results show that explanations of the MiniBooNE excess that involve $e^+e^-$ pairs from long-lived particles are in tension with T2K, PS191, and MicroBooNE data.
Similarly, these searches also constrain MiniBooNE explanations based on single photons due to the associated $e^+e^-$ decay mode via a virtual photon.
\end{abstract}

\maketitle

\section{Introduction}
The small nonzero neutrino masses challenges the conservation laws and particle content of the Standard Model (SM).
The existence of right-handed neutrinos $N_R$, singlets under the SM gauge symmetries, is an appealing solution to this puzzle.
In addition to participating in the Higgs mechanism, $N_R$ would admit a Majorana mass, possibly unrelated to the electroweak scale, and explain the neutrino masses via the seesaw mechanism~\cite{Minkowski:1977sc,*Mohapatra:1979ia,*GellMann:1980vs,*Yanagida:1979as,*Lazarides:1980nt,*Mohapatra:1980yp,*Schechter:1980gr,*Cheng:1980qt,*Foot:1988aq}.
In the simplest seesaw extensions, one is faced with a bleak experimental reality: either the heavy neutrino partners are too heavy to be produced in the laboratory, or their couplings to matter are too small to be observed.
However, this is not the case in low-scale variations of the model~\cite{Mohapatra:1986bd,*GonzalezGarcia:1988rw,Wyler:1982dd,*Akhmedov:1995ip,*Akhmedov:1995vm,Barry:2011wb,*Zhang:2011vh}, which are both ubiquitous and well-motivated theoretically, even if less predictive.
Among the most interesting cases is the inverse seesaw, where an approximate conservation of lepton number guarantees the smallness of neutrino masses in a technically natural way~\cite{naturalness_tHooft, Vissani:1997ys}.
This class of models predict the existence of (pseudo-)Dirac heavy neutral leptons (HNL) that can have mass below the electroweak scale and mix with SM neutrinos.

A HNL, $N$, interacts via the weak force suppressed by a small mixing element with SM neutrinos. 
This mixing is strongly constrained in the region between $\SI{10}\MeV$ and $m_K \simeq \SI{494}\MeV$~\cite{Atre:2009rg,deGouvea:2015euy,Drewes:2015iva,Fernandez-Martinez:2016lgt,Chun:2019nwi} thanks to laboratory-based searches, which provide upper bounds, and cosmological limits, which constrain the lifetime of $N$ to be $\tau_N < \mathcal{O}(0.1)\sec$, and therefore give a lower bound. 
These \emph{weaker-than-weak} interactions may not be the only contribution to their production or decays. 
Additional forces can modify significantly their decay widths even for couplings that would be otherwise very difficult to probe experimentally.
Of particular interest are scenarios wherein $N$ is shorter-lived than $\tau_N \lesssim \SI{0.1}\s$, so as to escape cosmological limits, but still sufficiently long-lived that it could survive $c\tau_N \gtrsim \SI{100}\m$, the typical distance from production to detection at beam-dump experiments.
These scenarios are most effectively constrained with these experiments, where $N$ could be copiously produced in meson decays and observed through its decay products inside large-volume detectors typically used for neutrino detection.

In this article, we consider decay-in-flight (DIF) searches at hodoscopic neutrino detectors for $N\to \nu e^+e^-$, and derive new bounds on the mixing between $N$ and muon-neutrinos.
Hodoscopic --- from the greek \textit{hodos} meaning path and \textit{scopos}, observer --- describes detectors that precisely reconstruct, track, and identify charged particles. 
This capability is essential for low-background searches for HNLs and other long-lived particles.
We consider three detectors: the T2K near detector ND280~\cite{T2KND280TPC:2010nnd}, MicroBooNE~\cite{MicroBooNE:2016pwy}, and PS191~\cite{Bernardi:1985ny,Bernardi:1987ek}.
We revisit constraints from PS191, thought to be the strongest, showing that they have been significantly overestimated.
We then extend the DIF search at T2K to HNLs lighter than the pion, showing that T2K data provides the leading lab-based constraints in that region of the minimal model.
These limits are then re-interpreted under three new scenarios with additional interactions between $N$ and the SM: a transition magnetic moment (TMM), a four-fermion leptonic interaction, and a leptophilic axion-like-particle ($\ell$ALP) portal.

We conclude by considering the latter two models in light of the excess of electron-like events in the MiniBooNE experiment~\cite{MiniBooNE:2018esg,MiniBooNE:2020pnu}.
In the case of a HNL with TMM interactions, we are able to partially cover the region of preference by constraining the off-shell photon decay into $e^+e^-$.
For the $\ell$ALP model, the HNLs decay predominantly into $e^+e^-$, and the limits we derive exclude that the MiniBooNE excess if fully explained by these events.
The code to obtain our limits and simulate HNL decays can be found on \textsc{g}it\textsc{h}ub.\footnote{\href{https://github.com/mhostert/}{{\large\color{purple}\faGithub}\,\,{github.com/mhostert/HNL-DIF}}.}

\section{Heavy Neutral Lepton Models}
\begin{figure}[t]
    \centering
    \includegraphics[width=\columnwidth]{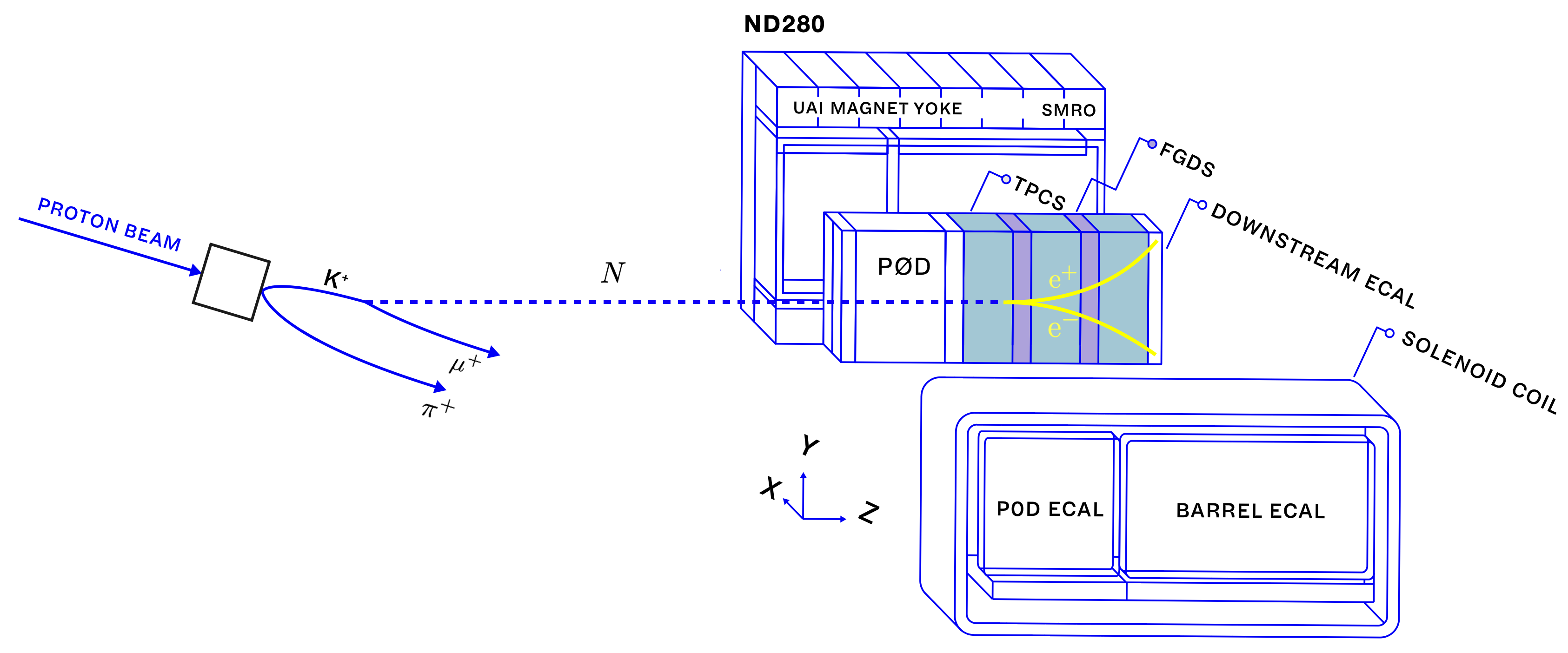}
    \caption{The T2K near detector, ND280, and the heavy neutrino decay-in-flight signature. The detector is located at an angle of $2.042^\circ$ with respect to the proton beam and at a distance of 284.9~m from the center of the production target.\label{fig:diagram}}
\end{figure}

\subsection{Minimal model}
The minimal model with a single HNL is defined by the low-energy Lagrangian
\begin{equation}
    -\mathscr{L_{\rm int}} \supset \frac{g}{2 c_W} U_{\alpha N}^* \overline{\nu_\alpha} \slashed{Z} P_L N + \frac{g}{\sqrt{2}} U_{\alpha N}^* \overline{\ell}_{\alpha} \slashed{W}  P_L N  + \text{ h.c.},
\end{equation}
where $N$ is the heavy mass eigenstate, which may be a Majorana or (pseudo-)Dirac particle, while $\alpha$ denotes any of the three SM flavors.
Although mixing with all three SM flavors is expected, we focus on dominant mixing with the muon-neutrinos, $|U_{e N}|, |U_{\tau N}| \ll |U_{\mu N}|$.
Our conclusions will be analogous in the case of dominant $|U_{e N}|$.
The case of dominant $|U_{\tau N}|$ is constrained by DIF searches at high-energy experiments, such as CHARM~\cite{Orloff:2002de} and NOMAD~\cite{NOMAD:2001eyx}, see also~\cite{Boiarska:2021yho}.
The weak decays of $N$ are straightforward to calculate, and we use the expressions in~\cite{Coloma:2020lgy}. 
Our work focuses on the decay $N \to \nu_\mu e^+e^-$ that proceeds via the neutral-current (NC) and we assume $N$ to be a Dirac particle. 

The long lifetimes of $N$ in the minimal model ($\tau^0 \sim \SI{1} \sec  \times (10^{-6}/|U_{\mu N}|^2)(100\text{ MeV}/m_N)^5$) has important consequences for cosmology.
In the early Universe $N$ will be thermally produced and, if it survives to the onset of Big Bang Nucleosynthesis (BBN), will impact the abundance of light elements~\cite{Sarkar:1995dd,Dolgov:2000jw,Ruchayskiy:2012si,Hufnagel:2017dgo}.
This happens in two ways: $N$ and its decay products upset the neutron-to-proton ratio, especially if $N$ can decay hadronically, in which case the known Helium abundance requires $\tau^0 < 0.023\sec$~\cite{Boyarsky:2020dzc}.
Moreover, its electromagnetic decay products heat up the plasma, changing the baryon-to-photon ratio, and impacting the deuterium abundance. 
Through-out this work, we use the detailed limits found in~\cite{Sabti:2020yrt}, neglecting effects from modified branching ratios.

\subsection{Non-minimal models}
Additional contributions to the HNL decay rate could make it decay before BBN.
We consider enhancements to the dilepton channel $N\to\nu e^+e^-$ from low-energy operators at dimension five and six, as well as from a low-energy extension with a light axion-like particle.

\paragraph{Transition magnetic moment} We start with the dimension-5, TMM operator
\begin{equation}\label{eq:dim-5}
    -\mathscr{L_{\rm int}} \supset \frac{\mu_{\rm tr}}{2} \overline{\nu_\alpha}\sigma^{\mu\nu}N F_{\mu\nu} + \text{h.c.},
\end{equation}
where motivated by phenomenological applications~\cite{Aparici:2009fh,Gninenko:2009ks,Gninenko:2010pr,Magill:2018jla,Shoemaker:2018vii,Brdar:2020quo,Vergani:2021tgc}, we set the flavor index $\alpha = \mu$.
If $|\mu_{\rm tr}| \gg (G_F m_N/ 2\sqrt{3}\pi)$, $N$ predominantly decays electromagnetically.
High-density detectors can observe the photon channel $N\to \nu_\mu \gamma$, while low-density detectors can measure the smaller rate to virtual photon, $\mathcal{B}(N\to \nu \gamma^* \to \nu e^+e^-)\sim 0.6\%$, benefiting from small neutrino-interaction backgrounds.


The operator in \cref{eq:dim-5} ought to be completed to restore $SU(2)_L$. 
This may bear consequence for the masses and mixing of neutral leptons, depending on the underlying model. 
If the completion of \cref{eq:dim-5} contains charged particles that couple only to heavy neutrinos, then $\mu_{\rm tr}$ can only be generated via mixing. 
In particular, a pseudo-Dirac pair of fermions $N_{L,R}$ with a large magnetic moment $(\mu_N/2) \overline{N_L}\sigma^{\mu\nu}N_RF_{\mu\nu}$ will generate a light-heavy transition moment in the mass basis, $\mu_{\rm tr} \sim U_{\alpha N}^* \mu_{N}$.
HNL decay rates are suppressed by mixing in this case.
On the other hand, if the new charged particle content couples to light neutrinos, then it may generate $\mu_{\rm tr}$ directly.
To avoid a relation between \cref{eq:dim-5} and the Dirac mass term $m_D\overline{\nu_L}N_R$ (and therefore to $U_{\alpha N}$), one may borrow several results from the literature on light Dirac neutrinos with large magnetic moments~\cite{Voloshin:1987qy,Barbieri:1988fh,Babu:1989px,Babu:1989wn,Leurer:1989hx}.
Following Voloshin's mechanism~\cite{Voloshin:1987qy}, for instance, if the completion of \cref{eq:dim-5} respects some approximate SU$(2)$ symmetry under which $(\nu_\mu, N_R^c)^T$ transforms as a doublet, then the dim-5 operator could be large and the mixing would be protected by the symmetry.
We assume that this is the case, rendering the lifetime of $N$ from \cref{eq:dim-5} independent of $|U_{\mu N}|^2$.

The TMM operator also generates a corresponding magnetic moment for $\nu_\mu$ due to the mixing $U_{\mu N}$. The $\nu_\mu-\nu_\mu$ magnetic moment is of the order of 
\begin{equation}
    |\mu_{\nu}| = |\mu_{tr} U_{\mu N}| \sim 3\times 10^{-11} \mu_B \times \left( \frac{|U_{\mu N}|}{10^{-2}} \right)\left( \frac{|\mu_{\rm tr}|}{1 \text{ PeV}^{-1}} \right)
\end{equation}
where $\mu_B$ is the Bohr magneton. This value is within the range of the XENON1T results~\cite{XENON:2020rca}, and therefore provides an upper limit on the mixing parameter in our plots.

\begin{figure}[t]
    \centering
    \includegraphics[page=1,width=\columnwidth]{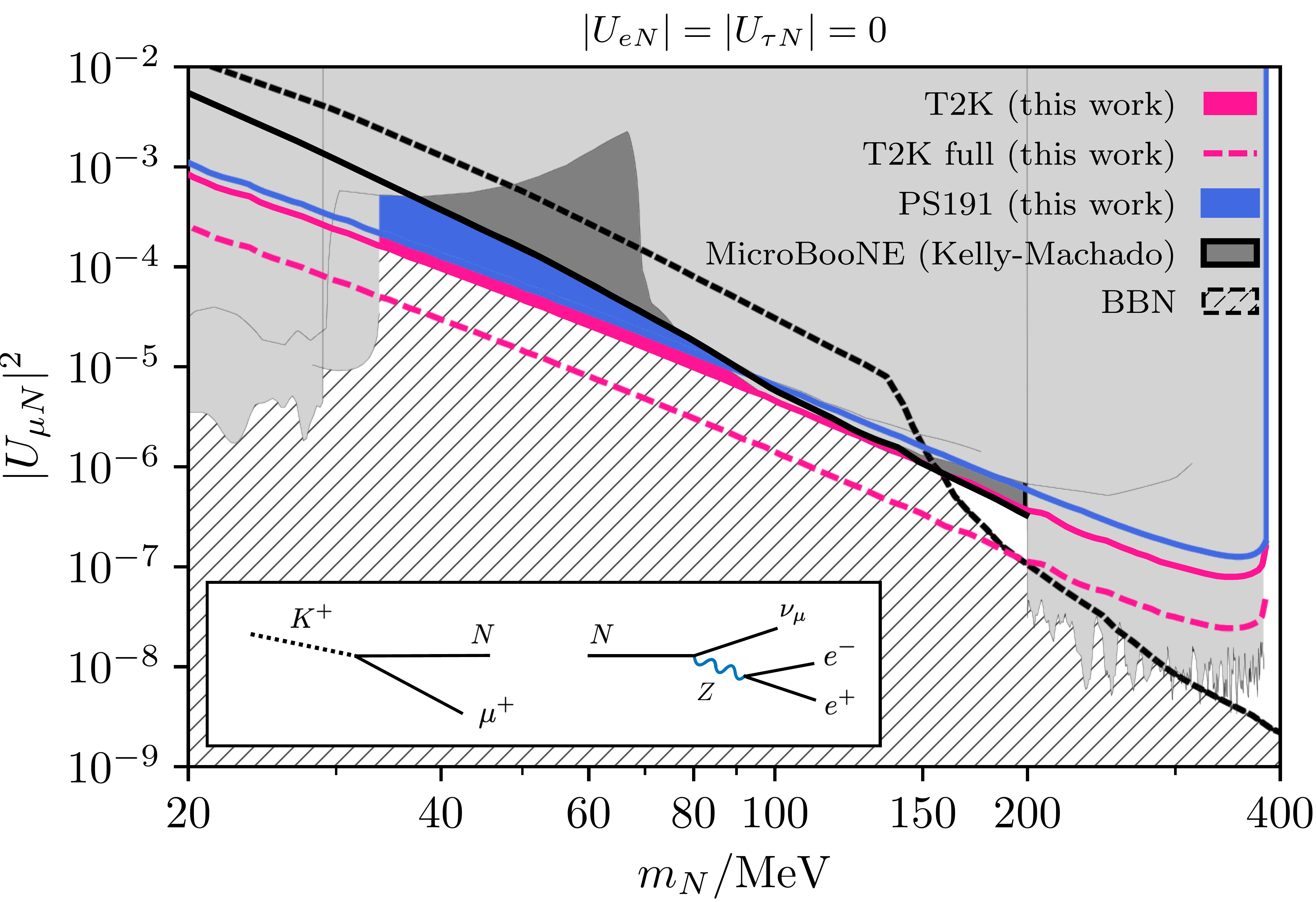}
    \caption{Constraints on the mixing of HNLs with the muon flavor as a function of its mass for a minimal HNL model at 90\% C.L. , considering only the production and decay mode: $K \rightarrow \nu_{\mu} N \rightarrow \nu_{\mu} (e^+e^-\bar{\nu_{\mu}})$.
    For MicroBooNE, T2K, and PS191 the regions above the lines are excluded, while BBN excludes the region below the line. In gray we show other model-independent constraints.
    T2K full refers to the projected sensitivity of T2K with the final dataset, which will be collected by the end of the experiment.
    \label{fig:minimal}}
\end{figure}
\begin{figure*}[t]
    \centering
    \includegraphics[page=2,width=0.49\textwidth]{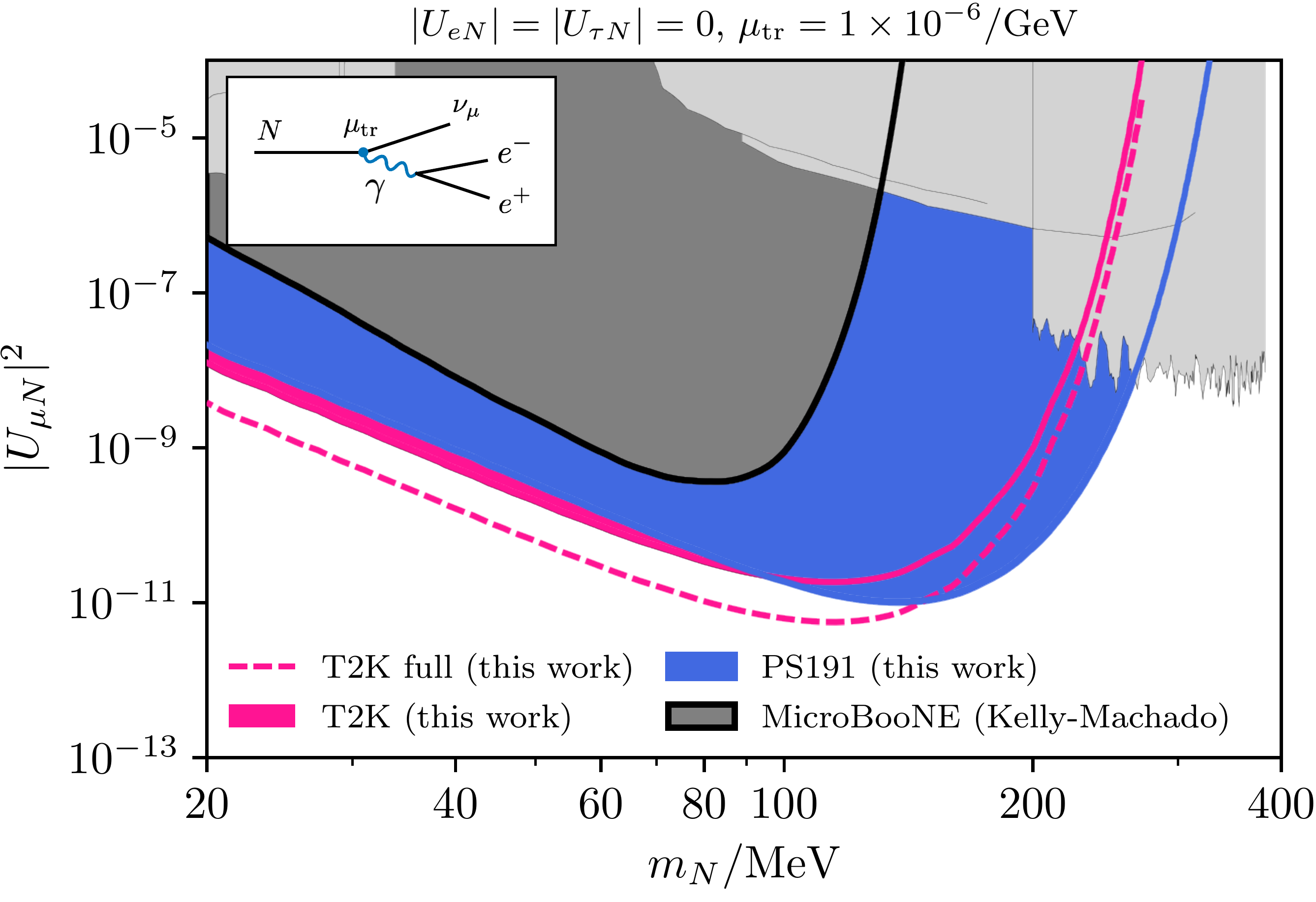}
    \includegraphics[page=3,width=0.49\textwidth]{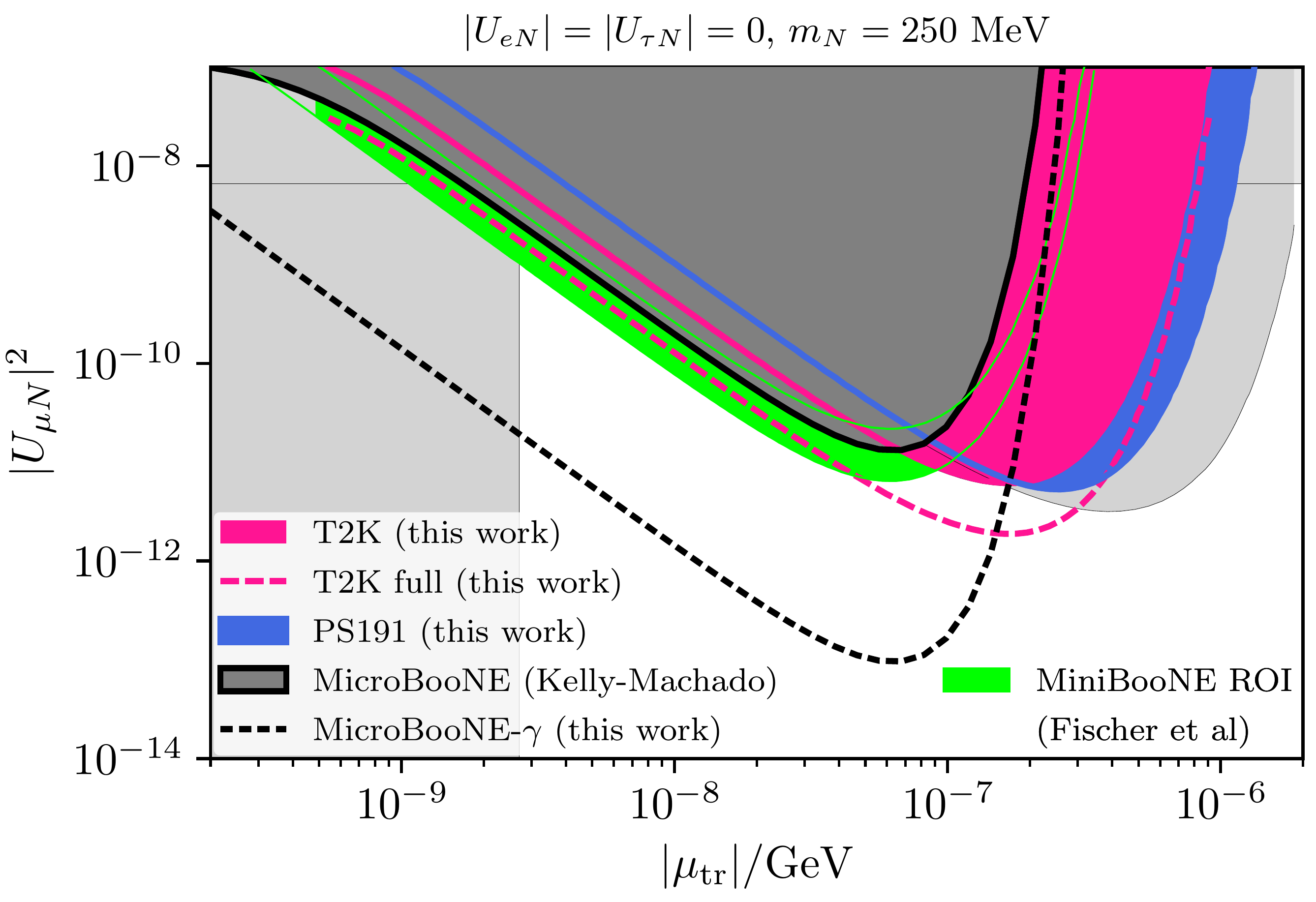}
    \caption{Left: Same as \cref{fig:minimal} but for HNLs with a TMM $\mu_{tr} = 1 \,\text{PeV}^{-1}$.
    Right: The same constraints as above but shown as a function of $\mu_{tr}$ for a fixed value of the HNL mass. The region of interest to explain the MiniBooNE excess is shown in green~\cite{Fischer:2019fbw}. 
    In fine dashed black we show an optimistic estimate of a NuMI-neutrino single-photon search at MicroBooNE.
    }
    \label{fig:dipole}
\end{figure*}

\paragraph{Four-fermion interaction} At dimension six, we consider a vectorial four-lepton interaction
\begin{equation}\label{eq:dim-6}
    -\mathscr{L_{\rm int}} \supset \frac{G_X}{\sqrt{2}} \left(\overline{N} \gamma^\mu N \right) \left( \overline{\ell_\beta} \gamma_\mu \ell_\beta \right) + \text{h.c.},
\end{equation}
where $\beta \in \{e,\mu,\tau\}$. For $G_F/G_X \ll 1$, HNLs decay primarily via $N \to \nu \ell_\beta^+ \ell_\beta^-$ through mixing with light neutrinos. 
Note that for large $G_X/G_F$, the effective operator in \cref{eq:dim-6} is only valid up to scales around the mediator mass. 
Therefore, constraints at high energies, for instance from the branching ratio of $Z\to N N (e^+e^-)$~\cite{Bilenky:1992xn}, do not apply. 

Ultraviolet completions of \cref{eq:dim-6} include gauge extensions of the Standard Model, such as $U(1)_{B-L}$ and dark $U(1)_X$ gauge symmetries~\cite{Bertuzzo:2018ftf,Abdullahi:2020nyr}.
In the latter, the dark photon $A^\prime$ couples to dark leptons, $g_X \overline{N_D} \slashed{A}^\prime N_D$, and to charged SM particles via kinetic mixing with the photon, $(\varepsilon/2) F_{\mu\nu} X^{\mu\nu}$. 
Thus \cref{eq:dim-6} is independent of flavor, resulting in \begin{equation}
    \frac{G_X}{\sqrt{2}} \sim \frac{g_X e \varepsilon}{m_{A^\prime}^2},
\end{equation}
where $g_X$ is the gauge coupling. 
As a result, the amplitude for $N\to\nu e^+e^-$ is proportional to $G_X U_{\mu N}$, shortening the HNL lifetime by a factor $\kappa \sim (G_F/G_X)^2$ with respect to the minimal model.
Model independent limits on kinetic mixing constrain $\kappa \gtrsim 10^{-4}$, and require the mediator to be relatively light, around the GeV scale. 
Above the dimuon threshold, $N\to \nu \mu^+\mu^-$ is allowed and would further strengthen our constraints.
In addition, for theoretical consistency, such hidden sectors typically contain several HNLs, which may also be produced in the decay of their heavier partners. 
Even faster decays like $N\to N^\prime \ell^+ \ell^-$ could dominate since the rate would be independent of $|U_{\mu N}|^2$. 
We do not comment further on this possibility, and assume that decay rates mediated by \cref{eq:dim-6} are always proportional to $|U_{\mu N}|^2$.

Finally, light neutrinos also interact via the dim-6 operator above due to mixing. There are two types of processes to consider: i) $\nu_\mu e \to N e$, and ii) $\nu_\mu e \to \nu_\mu e$. Process i) is kinematically forbidden for neutrino energies below $E_\nu^{\rm th} = m_N + m_N^2/2m_e$, which at the smallest masses we consider, $m_N = 20$~MeV, takes values of $E \sim 420$~MeV. This process would create a single electron shower inside experiments like MINER$\nu$A~\cite{MINERvA:2015nqi, MINERvA:2019hhc} and CHARM-II~\cite{CHARM-II:1994dzw}, which have measured the SM rate for neutrino-electron scattering at a precision of $\mathcal{O}(10\%)$ and $\mathcal{O}(3\%)$, respectively. Since signal i) comes from an inelastic scattering, the electron would be less forward and the signal efficiency due to stringent experimental cuts on $E_e \theta_{e}^2$ would be reduced. Nevertheless, requiring that the rate for process i) be less than $10\%$ of the weak rate and neglecting the effects of $m_N$, we find $|U_{\mu N}| G_X/G_F \lesssim 0.1$, which for our benchmark of $G_X/G_F =10^3$ gives $|U_{\mu N}| < 10^{-4}$. For process ii), the scattering on electrons is elastic, and therefore there is no threshold. Given that the new operator interferes with the SM amplitude, a naive scaling provides a limit of $|U_{\mu N}|^2 < 10^{-4}$ for our benchmark, again requiring the interference term to be below $10\%$ of the SM cross section. 

\paragraph{Light mediators} We now consider a low-energy extension of the SM where HNLs can decay to a light mediator, which in turn decays to $e^+e^-$. One simple example was already alluded to in the discussion above, where a dark photon mediator $X$ was proposed as a completion of the four-fermion interaction. While we focused on the case of $m_X > m_N$, justifying the effective operator approach, it may very well be possible that $X$ is lighter than $N$, so that it can be produced in two-body decays of the HNLs, $N\to \nu X$, to subsequently decay promptly into $e^+e^-$ via $X\to e^+e^-$. While a dark photon is attractive from a model-building perspective, it is certainly not the only one.

In Ref.~\cite{Chang:2021myh}, the authors proposed an extension of the SM with a leptophilic axion-like particle ($\ell$ALP). The simplified Lagrangian used was given by,
\begin{equation}\label{eq:ALPcouplings}
    -\mathcal{L} \supset \frac{\partial_\mu a}{2f_a}\left( c_N \overline{N}\gamma^\mu \gamma^5 N + c_e \overline{e}\gamma^\mu \gamma^5 e \right)
\end{equation}
where $f_a$ is the axion decay constant. The mixing of $N$ with active-neutrinos is then responsible for the decay of $N\to \nu a$, which overwhelms the branching ratios of $N$ for the parameter space of interest. The ALP decays promptly into $e^+e^-$ via the leptonic coupling. There are also loop-induced decays with $\mathcal{B}(a\to\gamma\gamma) \lesssim 10\%$. We account for these but do not consider it as part of our signal definition. 

The interactions in \refeq{eq:ALPcouplings} are again not gauge invariant and require a UV completion. As discussed in more detail in Ref.~\cite{Chang:2021myh}, the ALP $a$ can be identified with the pseudo-Goldstone boson of a global axial symmetry $U(1)_A$, under which $N$ is charged. It is also assumed that $a$ couples most strongly with electrons. While this assumption is not strictly necessary for our current study, it can be satisfied if the $U(1)_A$ symmetry has some non-trivial flavor structure, as, for example, in Froggatt-Nielsen models~\cite{Froggatt:1978nt}. 

Finally, as before, the new interactions also mediate neutrino-electron scattering. For $E_\nu \gg m_a, m_N$, the inelastic scattering cross section decreases with energy, 
\begin{align}
    &\sigma_{\nu_\mu e \to N e}(E_\nu \gg E_{\nu}^{\rm th})  \sim \frac{|U_{\mu N}|^2 |c_e c_N|^2 m_N^2 m_e^2}{256 \pi f_a^4} \frac{1}{2E_\nu m_e}
    \\
    & \quad \sim 8 \times 10^{-49} \text{ cm}^2\left(\frac{1 \text{ GeV}}{E_\nu}\right)\left(\frac{|U_{\mu N}|^2}{10^{-2}}\right)\left(\frac{100 \text{ GeV}}{f_a}\right)^4,\nonumber
\end{align}
for our benchmark of $m_N=20$~MeV, $c_N=0.4$ and $c_e = 1$, so safely below weak-interaction cross sections even for such large values of mixing. For $m_N$ values above $\mathcal{O}(100)$~MeV, the threshold becomes too large for $N$ to be produced in accelerator neutrino experiments. Elastic cross sections vanish in the limit of massless neutrinos.

\section{Decays in flight in neutrino detectors}

Accelerator neutrino beams are obtained from the DIF of magnetically-focused mesons. 
If HNLs exist, they are part of the neutrino beam produced through mixing.
The flux of HNLs in a given experiment can be estimated from the known neutrino flux per parent meson, by re-scaling it by~\cite{Shrock:1980vy,Shrock:1980ct}
\begin{equation}
\label{eqn:hnl_flux_scaling}
    \rho(a,b) = \frac{\Gamma_{M\to N \ell}}{\Gamma_{M\to \nu \ell}} = |U_{\ell 4}|^2 (a+b-(a-b)^2)\sqrt{\lambda(1,a,b)},
\end{equation}
where $M$ is the associated parent meson, $a=(m_\ell,m_M)^2$, $b=(m_N/m_K)^2$, and $\lambda$ is the K\"all\'en function. 
In this work, we include only the contribution from kaon decays, neglecting the additional production from pions and muons which would require a detailed experimental simulation. 
This procedure yields conservative results.
This rescaling method is sufficient in the low HNL mass region where $m_N \lesssim  (m_M - m_\ell)/2$, and automatically takes into account the effect of the magnetic focusing at the production point and other geometrical effects.
For larger $m_N$ values, our approach underestimates the HNL flux and therefore yields conservative results.
Heavier HNLs are produced with lower transverse momentum with respect to parent particles than light neutrinos. 
They are therefore more collimated with the beam direction, increasing the angular acceptance of on-axis detectors to HNLs, particularly at low energies where light neutrinos would have otherwise larger angular spread.

When considering new forces that shorten the lifetime, while keeping $N$ long-lived enough to reach the detector, the probability for $N$ to decay to some final state $X$ inside the detector is independent of the total lifetime ($1/\Gamma)$, and proportional only to the partial decay width in the signal channel ($\Gamma_{N\to X}$) 
\begin{align}\label{eq:prob_decay}
    P_{N\to X} &= e^{-L \Gamma/\beta\gamma} (1- e^{-\ell_{\rm det} \Gamma/\beta\gamma }) \, \mathcal{B}(N\to X) 
    \\\nonumber 
    &\simeq \frac{\ell_{\rm det}}{\gamma\beta} \, \Gamma_{N\to X},
\end{align}
where $\ell_d$ is the length of the detector, $L$ the distance between the production and decay points, $\gamma\beta = p_N/m_N$, and $\mathcal{B}$ denotes the branching ratio. 
Ultimately, in the long-lifetime and single-flavor-dominance limits, the new upper bound is given by $|U_{\alpha N}^{\rm new}|^2  = |U_{\alpha N}|^4  \times \widehat{\Gamma}_{N\to X} / \Gamma_{N\to X}^{\rm new}$, where $\widehat{\Gamma} = \Gamma/|U_{\alpha N}|^2$. 
The argument is analogous when relaxing the assumption of single-flavor dominance. 
If $\Gamma_{N\to X}^{\rm new} \propto |U_{\alpha N}^{\rm new}|^2$, the new upper-bound is proportional to the square-root of the ratio of signal decay rates, while the new lower bound, if it exists, will be linearly dependent on the ratio of the total rates. 
This is also approximately true for the lower bounds posed by cosmology.

\subsection{T2K ND280}The T2K collaboration searched for the DIF of HNLs in the three Gaseous Argon Time Projection Chambers (GArTPC) of the off-axis near detector ND280~\cite{Abe:2019kgx}.
Because of the low density of the argon gas, this search has very small backgrounds from neutrino interactions, while the gas allows excellent tracking and identification of the $e^+e^-$ final state.
The analysis observes no event in all channels, and provides some of the strongest limits in the mass region $\SI{140} \leq ~ m_N \leq \SI{493}\MeV$.
We use their null results and extrapolate the experimental efficiencies to estimate the constraint on light HNLs with $\SI{20} \leq ~ m_N \leq \SI{140} \MeV$.
We neglect systematic uncertainties and backgrounds, as they provide negligible contributions to the limits.
We reproduce the official T2K result above the pion mass with reasonable accuracy.

ND280 is currently being upgraded to a new configuration~\cite{T2K:2019bbb}, with the replacement of the $\pi^0$ detector, currently made of lead and scintillator, with two new GArTPCs.
DIF searches will benefit from the larger GAr volume and from the reduced number of backgrounds from coherent neutrino interactions upstream of the TPCs.
We estimate the sensitivity of a future search with this upgrade by considering the increased volume, and a total of $2\times 10^{22}$ POT~\cite{Abe:2016tii}, $4\times 10^{21}$ before (already collected), and $16\times 10^{21}$ after the upgrade.
This is a conservative estimate that neglects improvements to reconstruction and background rejection.

\subsection{MicroBooNE}
MicroBooNE is the first ton-scale liquid argon (LAr) TPC operated in a neutrino beam.
It can perform searches for the DIF of new particles~\cite{Batell:2009di,Ballett:2016opr,Batell:2019nwo}.
However, since LAr is a high-density material, providing both the target and the detector material, one needs to rely on extra schemes to reject neutrino-induced background.
The delayed arrival of HNLs with respect to neutrinos 
was explored in the search for HNLs in the $\mu\pi$ decay channel~\cite{microboone_hnl}.
The authors of~\cite{Kelly:2021xbv} recast the null result of the MicroBooNE~\cite{MicroBooNE:2021usw} search for light scalars into limits on the HNL mixing.
This search successfully rejects background utilizing the directionality and time of arrival of new particles, as produced in decays at rest of kaons at the beam dump of the NuMI beam.
We further re-interpret this constraint as bounds on the non-minimal models considered.
These results use the reconstruction described in~\cite{MicroBooNE:2017xvs} and could be improved in MicroBooNE and future LArTPC by using new and innovative reconstruction techniques~\cite{MicroBooNE:2020yze, MicroBooNE:2020hho, MicroBooNE:2021ddy,  MicroBooNE:2021ojx}.
Additionally, MicroBooNE performed searches for signatures that could explain the MiniBooNE excess in terms of electron \cite{MicroBooNE:2021rmx, MicroBooNE:2021sne, MicroBooNE:2021jwr, MicroBooNE:2021nxr} and single photon \cite{MicroBooNE:2021zai} production in the detector.
Although these results do not report excesses with respect to the standard model expectation, they do not fully exclude that the MiniBooNE anomaly is related to new particles in the beam, as discussed in \cite{Arguelles:2021meu}.
Moreover, these analyses do not significantly constrain the models we are considering in this paper, because they do not include a tailored search for $e^+e^-$ from decay in flight, while the search for single photons with no protons in the final state is not sensitive to event rates compatible with the MiniBooNE excess.

\begin{figure}[t]
    \centering
    \includegraphics[page=4,width=\columnwidth]{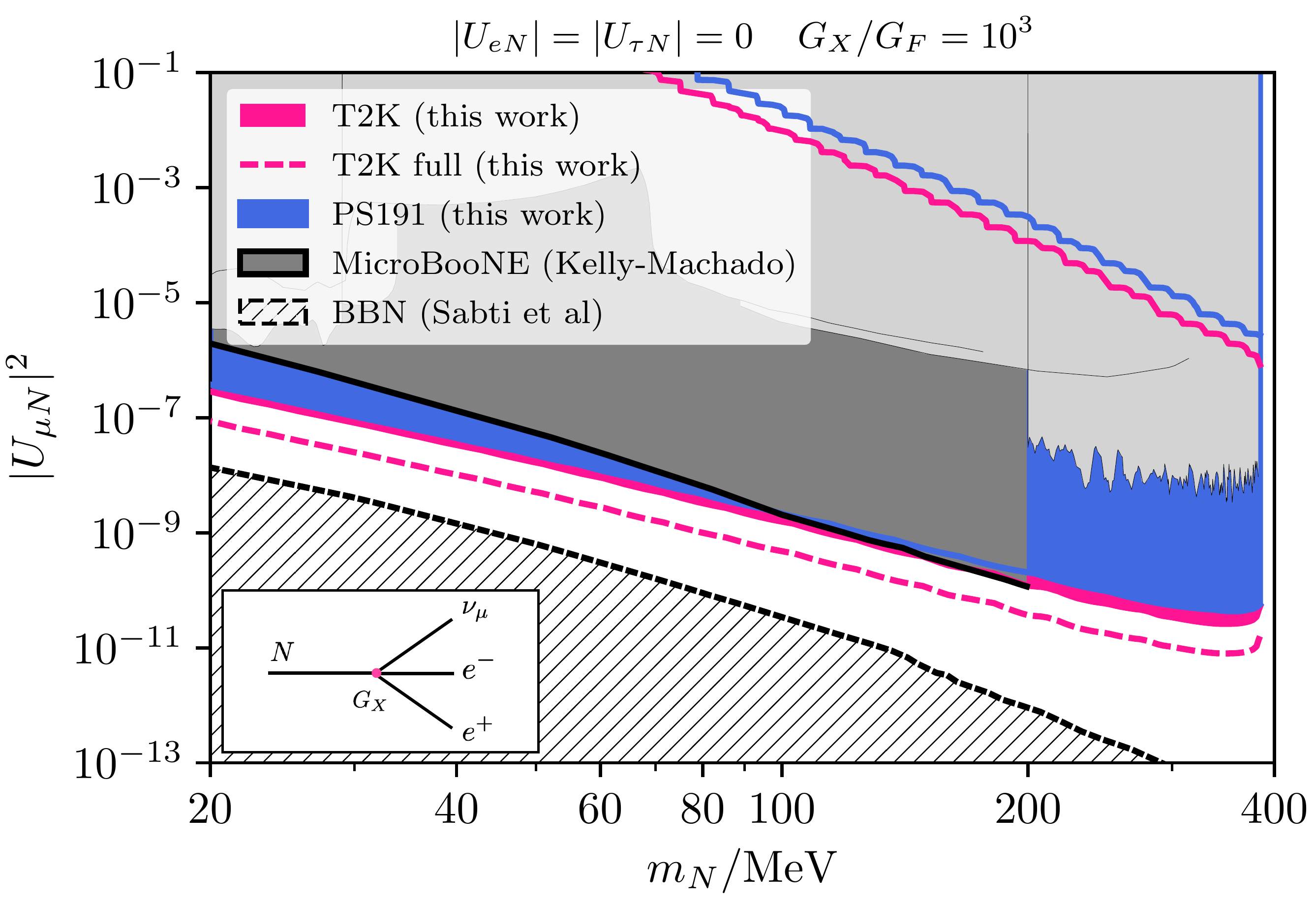}
    \caption{Same as \cref{fig:minimal} but for HNLs with a four-fermion leptonic interaction with $G_X = 10^{3} G_F$. The inset diagram shows the dominant decay process in the model.}
    \label{fig:hidden_sector}
\end{figure}

\begin{figure*}[t]
    \centering
    \includegraphics[page=5,width=\columnwidth]{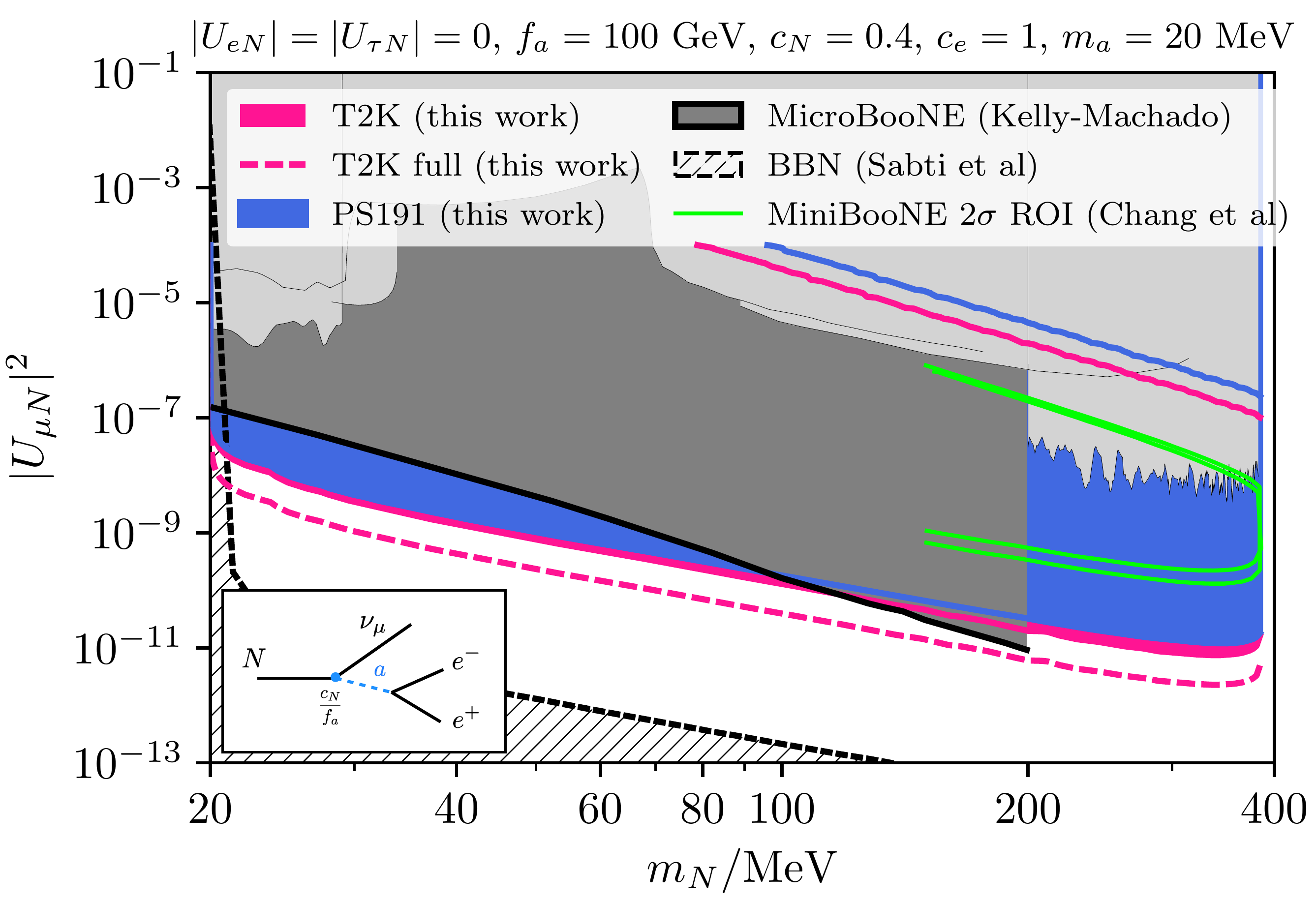}
    \includegraphics[page=6,width=\columnwidth]{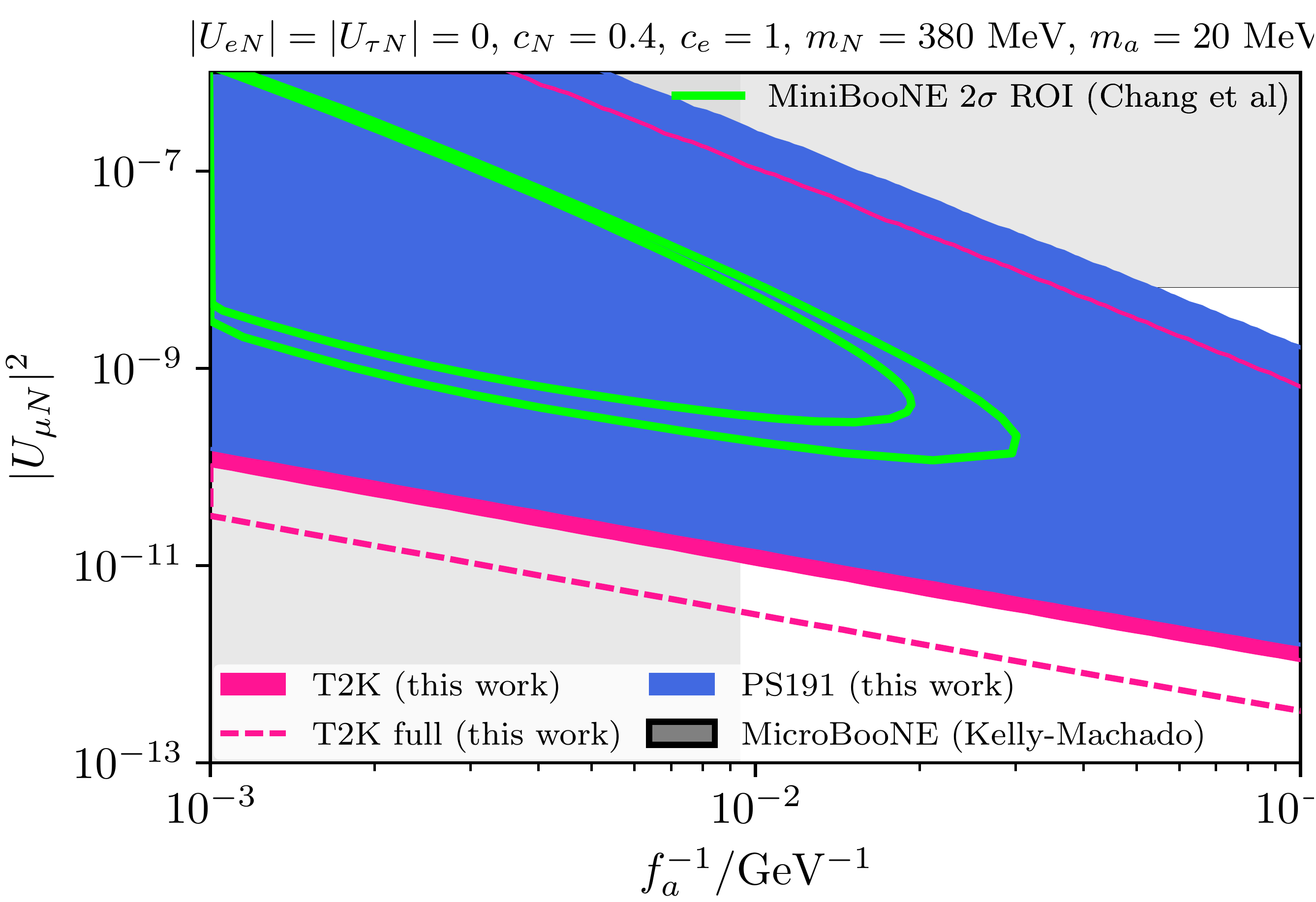}
    \caption{Same as \cref{fig:minimal} but for HNLs that decay to a light leptophilic ALP of $m_a = 20$~MeV. The inset in the left panel illustrates the dominant decay mode of the HNLs in this model. The MiniBooNE region of preference at $2\sigma$ is defined by the two solid green lines.}
    \label{fig:alp}
\end{figure*}

\subsection{PS191}PS191 was a low-density detector located at CERN and designed to search for the decays of long-lived particles. 
The 19.2~GeV proton beam provided a neutrino beam for the on-axis BEBC experiment, and could also be used by PS191 at 40~mrad off-axis location, where $\langle E_\nu \rangle\sim 1$~GeV.
The detector comprised of helium bags separated by scintillator planes followed by a dense electromagnetic calorimeter downstream~\footnote{The calorimeter was used in a search for $\nu_\mu \to \nu_e$ oscillations, where an excess was observed~\cite{Bernardi:1986hs}.}. 
Their final results considered only the CC decays of Dirac HNLs~\cite{Levy:1986yx,Bernardi:1987ek}.
~\footnote{In the first publication~\cite{Bernardi:1985ny} it was incorrectly stated that the limits are independent of the Dirac or Majorana nature of the HNLs, which was later corrected in~\refrefs{Levy:1986yx}{Bernardi:1987ek}.}. 
As a consequence, the search for $K^+\to\mu^+ (N \to \pi^+\mu^-)$ and $K^+\to \mu^+ (N\to \nu_e \mu^- e^+)$ was used to constrain only the $|U_{\mu N}|^2$-dominant case~\footnote{We note that the channel $K^+ \to \mu^+ (N \to \nu_\mu \mu^+ \mu^-)$ was not considered even though it also proceeds via CC diagrams and could also constrain $|U_{\mu N}|^2$.} and the search $K^+\to\mu^+ (N \to e^+e^-)$ was used to constrain $|U_{\mu N} U_{e N}|$. 
We are concerned with the latter case, as even in a $|U_{\mu N}|^2$-dominant case, HNLs still decay to $\nu e^+e^-$ via NC and a constraint can be derived. 
This point was first appreciated in~\cite{Kusenko:2004qc} and later discussed in~\cite{Ruchayskiy:2011aa,Drewes:2015iva}. 
These constraints were thought to be the strongest lab-based ones in this mass region for the $|U_{\mu N}|^2$-dominant case. 
With our simulation, we show that this is not the case.
The bound on $|U_{\mu N}|^2$ is a factor of 6 weaker than the published ones, corresponding to an event rate 36 times smaller~\footnote{We have found a similar factor in the channels $N\to \mu \pi$ and $N\to \nu e \mu$. For the latter, we find a discrepancy by a factor of 6 at the largest HNL masses, which decreases for lower HNL masses. We attribute this effect to us overestimating the signal efficiency. For $\mu\pi$, the discrepancy is a constant factor of 7.5.}. 
This is corroborated by the fact that T2K and PS191 have very similar total exposures and neutrino fluxes, noting that PS191 ran for only a month.
(Supplemental Material ~\footnote{See Supplemental Material at - URL will be inserted by publisher - for a comparison between PS191 and T2K, as well as more information about the HNL decay rate.})

If constraints on light scalars can be translated to constraints on HNLs, as illustrated in~\cite{Kelly:2021xbv}, it follows that the converse is also true. For example, \cite{Gorbunov:2021ccu} used the null results from PS191 to cast limits on the light scalar $\phi$. 
For the muon-dominant case, this is a trivial translation, as $K\to \pi \phi$ and $K\to \mu N$ have very similar kinematics. 
Our new PS191 limits recasted for scalars agree with~\cite{Gorbunov:2021ccu}. 
We also note that recasting our results below for current (future) T2K data, we find a constraint on the scalar mixing of $\theta< 2.3 \, (1.5) \times 10^{-4}$ for $m_\phi = 150$~MeV, which is the leading constraint in the ``pion gap''~\cite{Fuyuto:2014cya}.

\section{Results and discussion} 
Our results for the minimal model are shown in \cref{fig:minimal}. 
Our bounds and sensitivity estimates rule out HNLs below the kaon mass with dominant muon mixing in the minimal model. 
This result also puts more strain on models that could also account for the baryon asymmetry of the Universe~\cite{Bondarenko:2021cpc}.
We have also shown that previous limits from PS191 have been overestimated by an order of magnitude, and that T2K provides the leading constraints at the lowest masses. 
Future data can improve the leading constraints below $\SI{200}\MeV$, where the kaon peak searches become insensitive.
We note that our results are based on extrapolated efficiencies and conservative flux simulation, and that a complete simulation within the collaboration will provide improved results.

The new constraints with decays via the dimension five, \cref{eq:dim-5}, are shown in \cref{fig:dipole}, while for dimension six, \cref{eq:dim-6}, in \cref{fig:hidden_sector}. 
In these scenarios the combination of lab-based and cosmological constraints do not exclude HNLs below the kaon mass.
Our work complements searches for neutrino upscattering at CHARM-II, which provides stronger constraints for new physics scales below $\SI{1}\PeV$~\cite{Coloma:2017ppo,Magill:2018jla}, and constraints from supernova, which dominate above $\sim\SI{1}\EeV$.
For $G_X/G_F=10^{3}$, BBN constraints still exclude the smallest mixings, but our lab-based results provide the best upper limits in the newly allowed parameter space. 
For this choice of parameters, one expects the existence of a new vector mediator with a mass of a few GeV, which can be searched for in collider experiments~\cite{Fabbrichesi:2020wbt}.

Our constraints are relevant to new physics explanations of the MiniBooNE excess of electron-like events~\cite{MiniBooNE:2018esg,MiniBooNE:2020pnu}. 
The authors of~\cite{Fischer:2019fbw} proposed that the excess can be explained by the DIF of HNLs with a TMM. 
This observation can be generalized to any model with enhanced $N\to \nu e^+e^-$ or $\nu \gamma$ rates. 
The MiniBooNE region of interest and our constraints in~\cref{fig:dipole} show that MicroBooNE and T2K are already constraining interesting parameter space for the TMM model.
Future T2K data will shine further light on this interpretation.
Additionally, future iterations of the MicroBooNE analysis will benefit from a larger dataset, about ten times more statistics, and from new and improved reconstruction and analysis techniques, such as the WireCell \cite{MicroBooNE:2021ojx} and the Deep Learning \cite{MicroBooNE:2020yze} frameworks.

Constraints on the HNL decays to $\ell$ALP are shown in \cref{fig:alp}. On the left panel we show the limits in mass and mixing for fixed values of the HNL-$\ell$ALP couplings and $\ell$ALP mass. On the right, we fix the HNL mass to be $380$~MeV and vary the HNL mixing and the $\ell$ALP decay constant, $f_a$. As it can be seen, the limits from T2K and PS191 fully cover the region of preference to explain the MiniBooNE excess in both panels. On the right panel, we do not show the MicroBooNE limits as the HNL mass is beyond the range considered by the experiment. However, inspecting the left panel, one can deduce that MicroBooNE would also strongly constrain the large mass region of the explanation. Similarly to the TMM model, the invariant mass of the $e^+e^-$ produced in HNL decays is very small, in fact, $m_{ee} = m_{a}$. For MicroBooNE, separating the lepton pair would be more challenging due to the absence of magnetic fields. When setting our limits in this model, we assign the final signal efficiency to be that of a standard HNL model with $m_N = m_a$, that is, we set $\epsilon_{\ell {\rm ALP}} = \epsilon(m_N = m_a)$.

We now comment on the kinematics of HNL decays. 
The NC and CC contributions to $N\to \nu e^+e^-$ decays, as well as their interference, do not differ significantly in regards to the experimental variables upon which the signal efficiency depends.
This includes the distributions of the $e^+e^-$ separation angle and their individual energies.
This remains true for both the Dirac and Majorana cases, as well as for the additional interactions in \cref{fig:dipole} and \cref{fig:hidden_sector}, rendering the constraints mostly independent of the specifics of the HNL interactions. 
The most distinct case is the TMM, where the final products have smaller opening angles and are less collimated with the beam.
Nevertheless, in most events the physical separation between the electrons remains large, especially with magnetic fields.
The impact of the new interactions on the selection efficiency should be studied by the T2K collaboration with its full detector simulation, but we do not expect significant changes to our results.

To illustrate the impact of making use of the $e^+e^-$ branching ratio instead of the single-photon rate, we also draw on on \cref{fig:dipole} the sensitivity estimate of a search for single photons with the same efficiencies, backgrounds, and exposure as the $e^+e^-$ search at MicroBooNE.
This complementary strategy can only be achieved in high-density detectors, but is limited by the neutrino-induced backgrounds.  
With a volume of 0.66 and 4.5 times the \muboone~volume, respectively, and a similar beam exposure, SBND~\cite{sbnd} and ICARUS~\cite{icarus} could complement the MicroBooNE constraints, thanks to the different distances from beam target and absorber.
While this projection relies on optimistic assumptions on the reconstruction and selection efficiency for a photon converting in the detector, which are likely to be lower than for a genuine $e^+e^-$ pair and have higher backgrounds, even if the rate estimation would higher by a factor of ten, lowermost of the MiniBooNE preferred region will be tested.

When the HNLs are too short-lived to be probed in DIF searches, they may be produced by coherent neutrino-nucleus upscattering in the dense lead layers of ND280~\cite{Gninenko:2009ks,Gninenko:2010pr,Coloma:2017ppo,Vergani:2021tgc,Bertuzzo:2018itn,Ballett:2018ynz,Ballett:2019pyw}.
A detailed study of this scenario is in progress~\cite{upcoming}. 
Prompt HNL decays in pion and kaon factories, such as PIENU~\cite{PIENU:2017wbj,PIENU:2019usb} and NA62~\cite{NA62:2021bji,NA62:2020mcv}, should also be searched for. 
The channel $K^+\to\ell^+ (N \to \nu e^+e^-)$, proposed in~\cite{Ballett:2019pyw}, would be sensitive to light dark sector models and, to a lesser extent, to TMM.

\section{Conclusions}

We have studied a heavy neutral lepton (HNL) extension of the Standard Model (SM), focusing on the case where HNLs mix predominantly with the muon flavor. In a minimal HNL model, we have derived new limits using T2K data at HNL masses below the pion. We do so by means of a linear fit to the signal efficiencies, extrapolated down to the lowest masses. This approach can be improved upon by collaboration, but we note that at low masses, the signal efficiency should not vanish due to the magnetic field inside the ND280 detector. We also revisit the PS191 limits on HNLs, and find a large discrepancy with the published limits, weakening the previously-published limits by, approximately, a factor of 36 in the event rate.

We have also shown that additional interactions in a heavy neutrino sector can reconcile HNLs lighter than the kaon with cosmology. New portal interactions between the HNLs and the Standard Model can sufficiently decrease the HNL lifetime so as to deplete their number density by Big-Bang-Nucleosynthesis, while keeping it long enough to allow for decay-in-flight signatures in neutrino experiments. We have considered a transition magnetic moment portal, a four-fermion interaction with electrons, such as that generated by a GeV-scale dark photon, and a light leptophilic axion-like particle model. In all cases, new parameter space for $N\to \nu e^+e^-$ decay-in-flight signatures opens up for masses below the kaon mass. 

For the transition magnetic moment and the axion-like particle models, the proposed solutions to the MiniBooNE excess are constrained by our limits. In the former model, our limits are derived using off-shell photon decays, therefore suppressing the $e^+e^-$ signal rate at T2K, PS191, and MicroBooNE with respect to the $\gamma$ signal rate at MiniBooNE by approximately $\alpha/4/\pi \times \log{(m_N^2/m_e^2)} \approx 6\times 10^{-3}$ in the parameter space of interest. For the axion-like particle, we find that T2K and PS191 fully exclude the MiniBooNE region of preference by over an order of magnitude, excluding that the excess is due to decays in flight to a great significance.

Low as well as high density hodoscopic detectors like ND280 and MicroBooNE can play a central role in the search for long-lived particles.
Due to its hybrid design, ND280 can place strong limits on upscattering production of light particles, like dark neutrinos~\cite{Bertuzzo:2018itn,Ballett:2018ynz,Ballett:2019pyw,upcoming} and co-annihilating dark matter~\cite{Tucker-Smith:2001myb,Izaguirre:2017bqb}.
Future detectors, such as the planned DUNE near detector, could also benefit from a hybrid detector design, since they would be sensitive to both charged-track and single photon final states, while having a region of low neutrino-induced backgrounds.

\acknowledgments
We acknowledge useful discussions with Kevin Kelly, Mathieu Lamoureux, Pedro Machado, Sophie King, and Teppei Katori.
C.A.A. is supported by the Faculty of Arts and Sciences of Harvard University, and the Alfred P. Sloan Foundation. 
N.F.’s work is supported by the Department of Energy grant award number DE-SC0007881.
The research of M.H. was supported in part by Perimeter Institute for Theoretical Physics. Research at Perimeter Institute is supported by the Government of Canada through the Department of Innovation, Science and Economic Development and by the Province of Ontario through the Ministry of Research, Innovation and Science. Part of M.H.'s work was performed at the Aspen Center for Physics, which is supported by National Science Foundation grant PHY-1607611.

\appendix

\section{Decay rates}
\label{app:kinematics}

All models we discuss predict HNL decays into $e^+e^-$ final states. In this section we discuss how the experimental signatures may differ under different model hypotheses. We compare the double differential decay rate among all three models considered, also paying attention to the Dirac and Majorana distinction. It will become clear that the efficiencies are not expected to differ substantially between different models, even if one can find variables where the distributions do differ significantly. For a full discussion regarding the difference between Dirac and Majorana HNLs, see~\cite{deGouvea:2021ual}.

\begin{figure*}[h]
    \centering
    \includegraphics[width=0.32\textwidth]{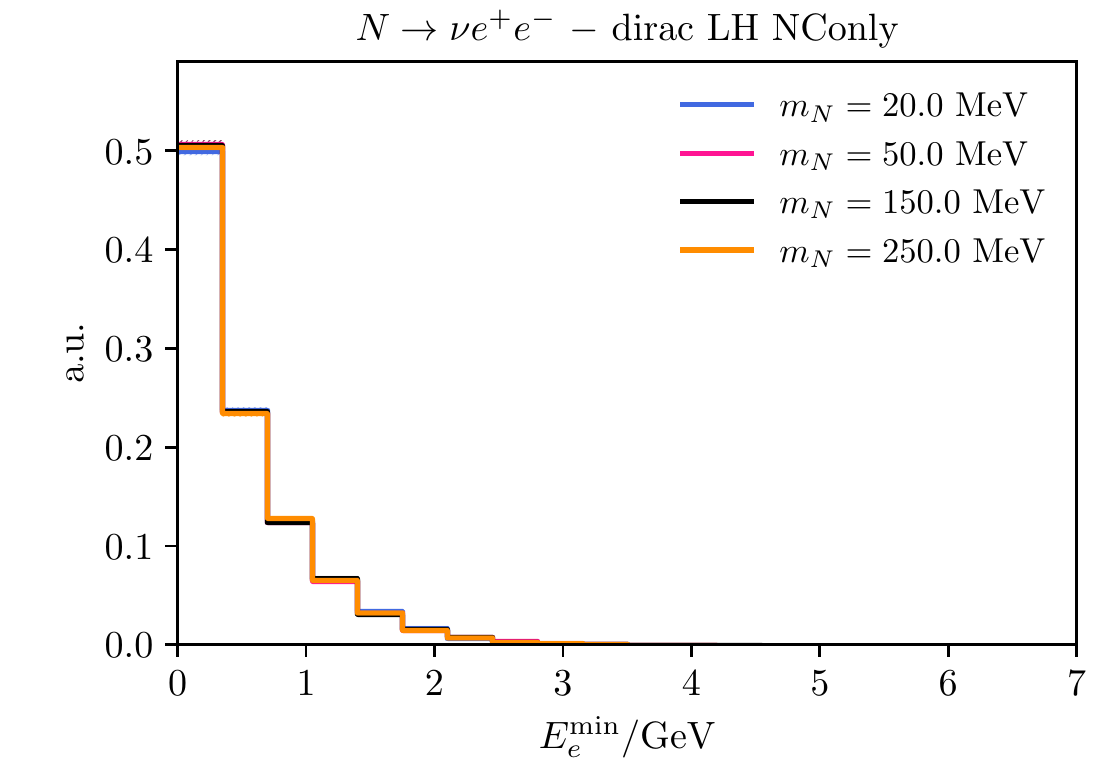}
    \includegraphics[width=0.32\textwidth]{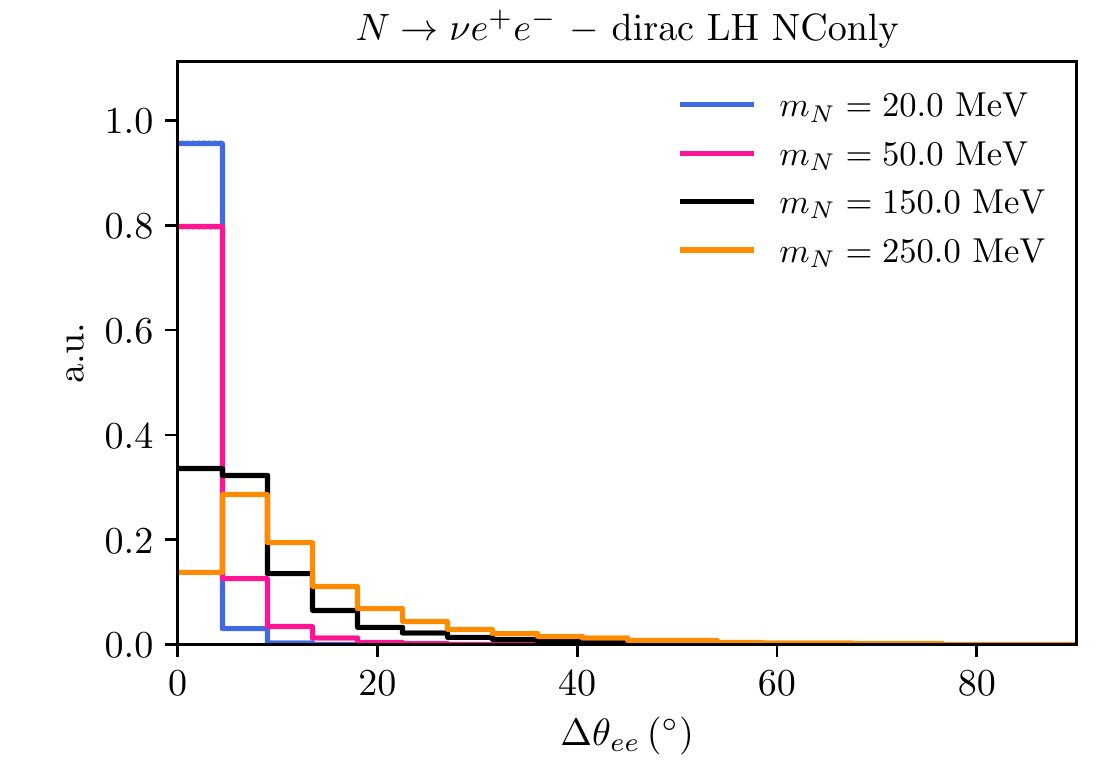}
    \includegraphics[width=0.32\textwidth]{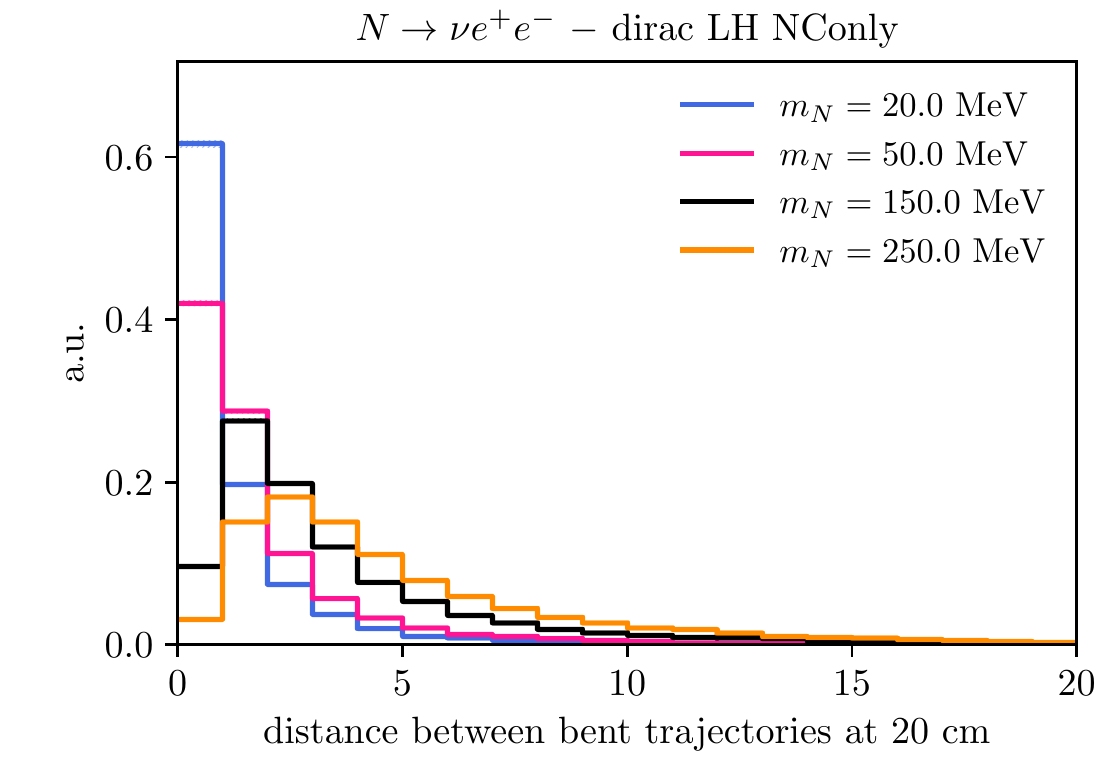}    
    \caption{Kinematic distributions for several HNL masses. We assume the $N\to \nu e^+e^-$ decay proceeds purely via neutral current (NC) diagrams. Left: the energy of the lowest energy particle of the electron-positron pair. Center: the opening angle between the $e^+e^-$ pair. Right: the distance between $e^+e^-$ pair at $\SI{20}\cm$ from the decay point along the direction of the total $e^+e^-$ momentum. We take into account the bent trajectories of pair assuming a constant $B=\SI{0.2}\T$ magnetic field. Monte-Carlo errors are too small to be seen. \label{fig:distribution_masses}}
\end{figure*}

\begin{figure*}[h]
    \centering
    \includegraphics[width=0.32\textwidth]{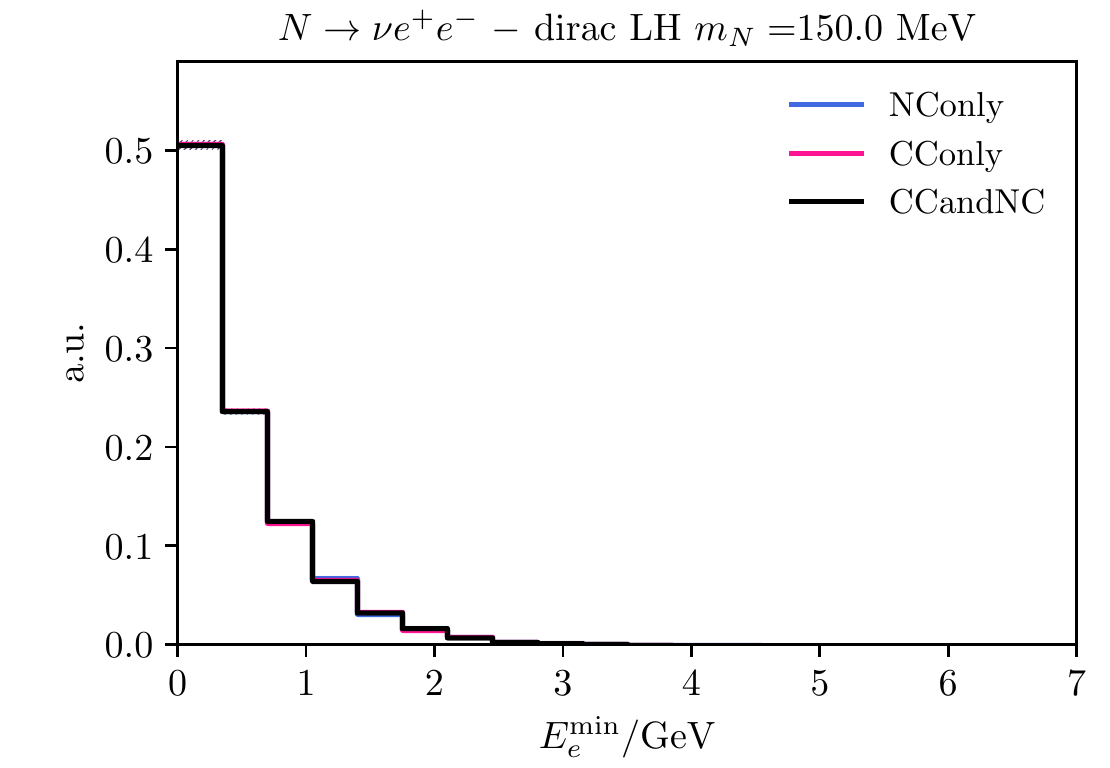}
    \includegraphics[width=0.32\textwidth]{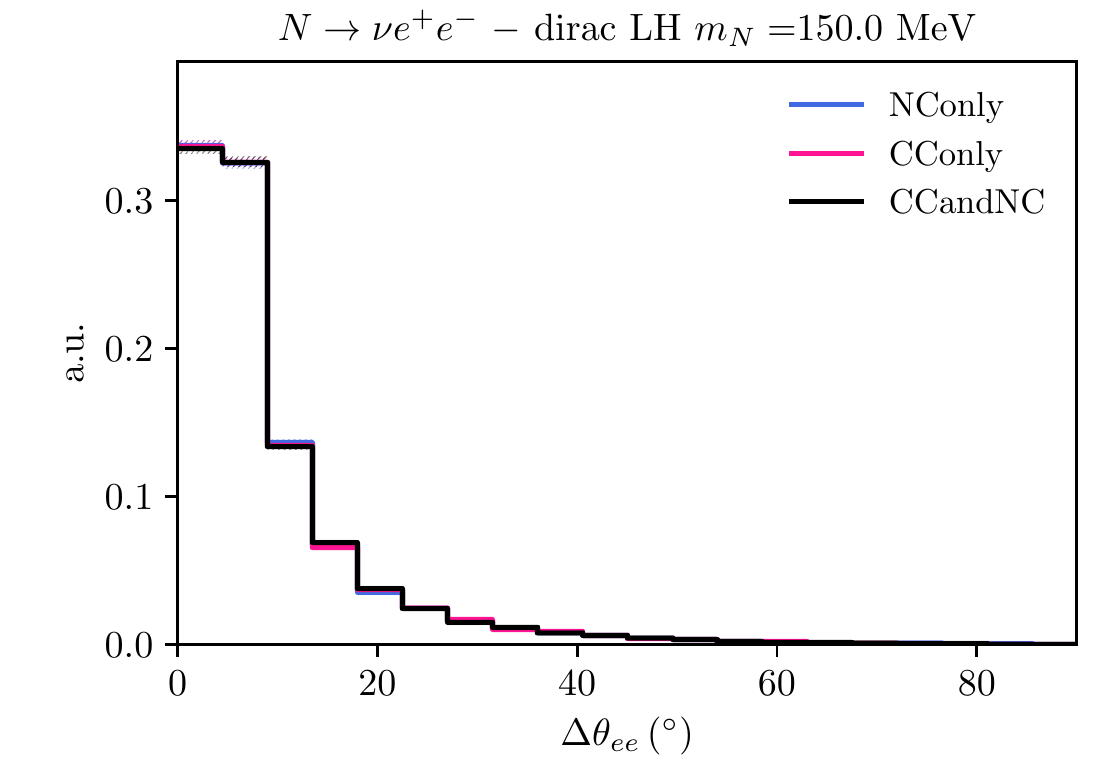}
    \includegraphics[width=0.32\textwidth]{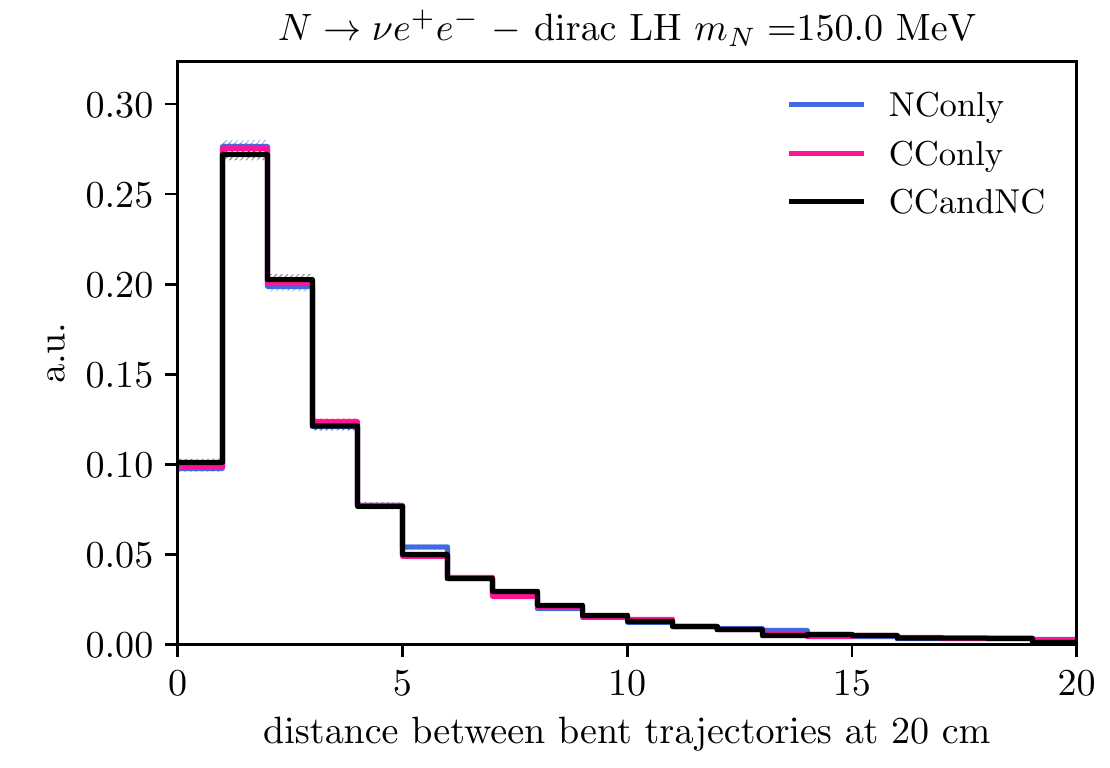}    
    \caption{Same as \cref{fig:distribution_masses} but varying the diagram leading to $N\to \nu e^+e^-$. The kinematics is mostly independent of the nature of the decay. \label{fig:distribution_vertices}}
\end{figure*}

\begin{figure*}[h]
    \centering
    \includegraphics[width=0.32\textwidth]{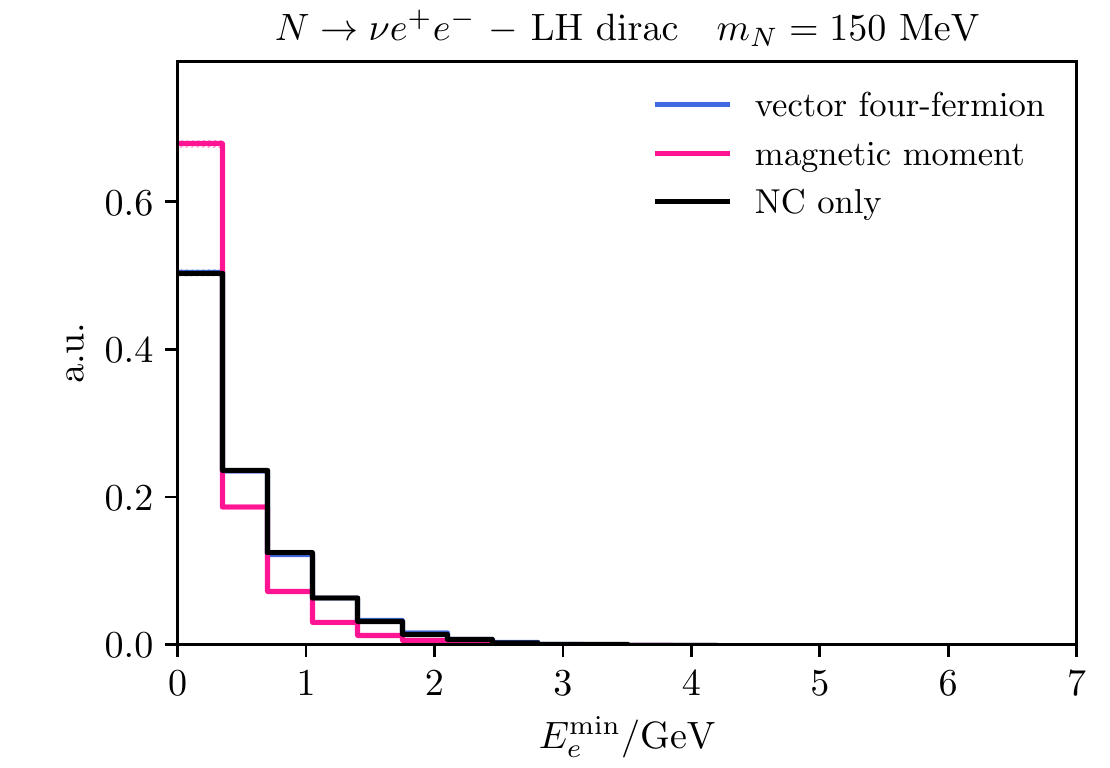}
    \includegraphics[width=0.32\textwidth]{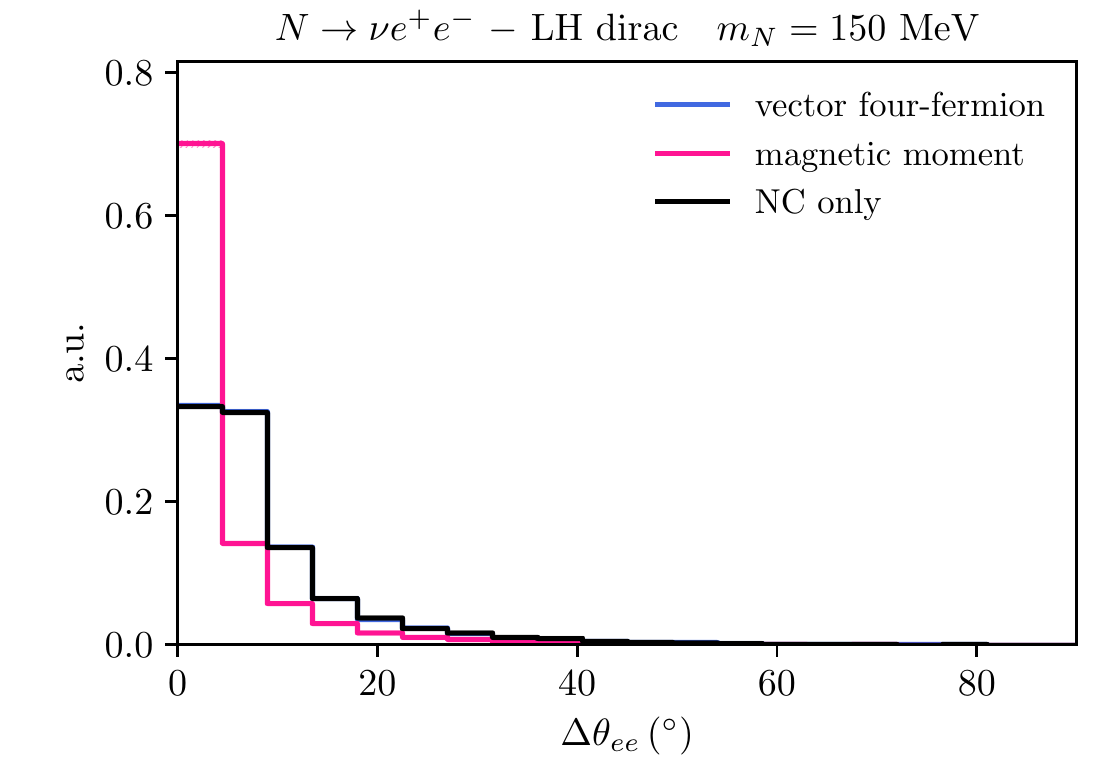}
    \includegraphics[width=0.32\textwidth]{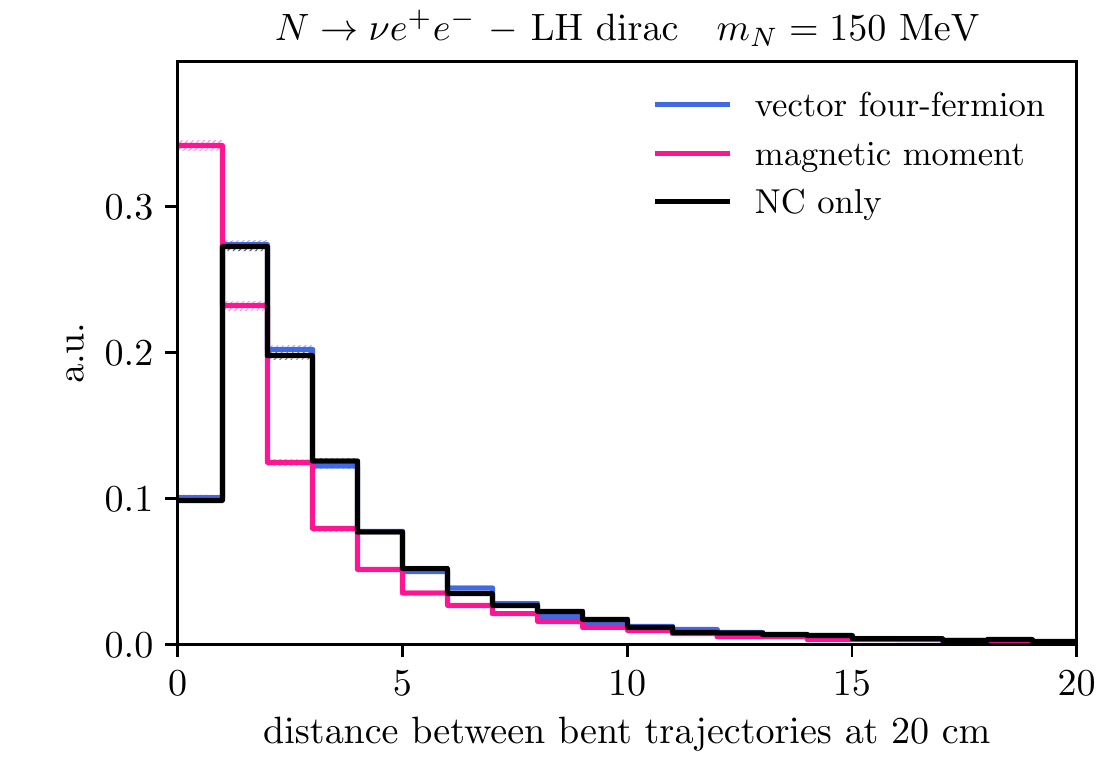}
    \\
    \includegraphics[width=0.32\textwidth]{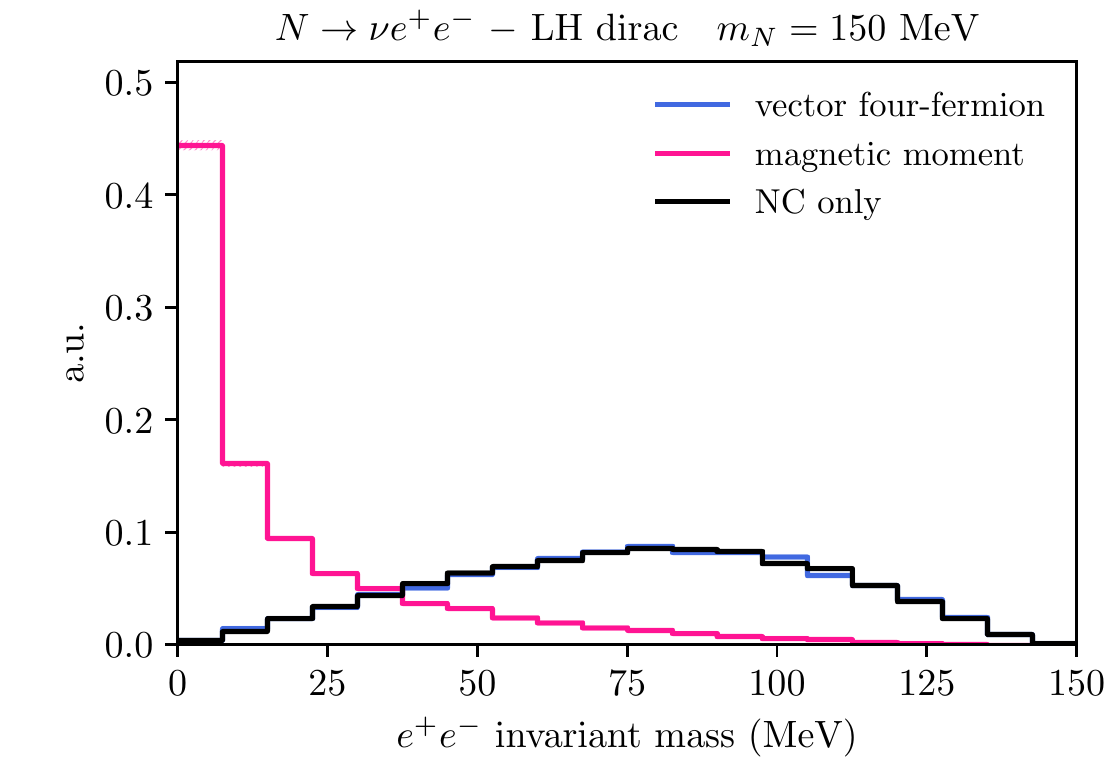}
    \includegraphics[width=0.32\textwidth]{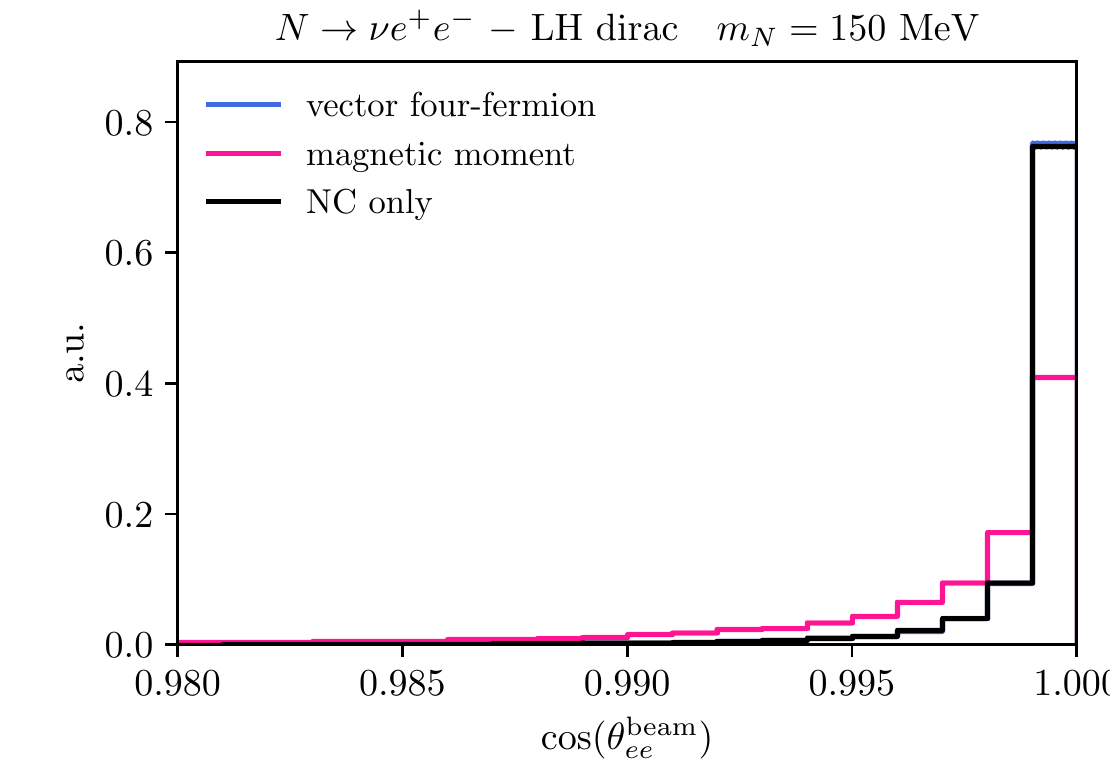}
    \includegraphics[width=0.32\textwidth]{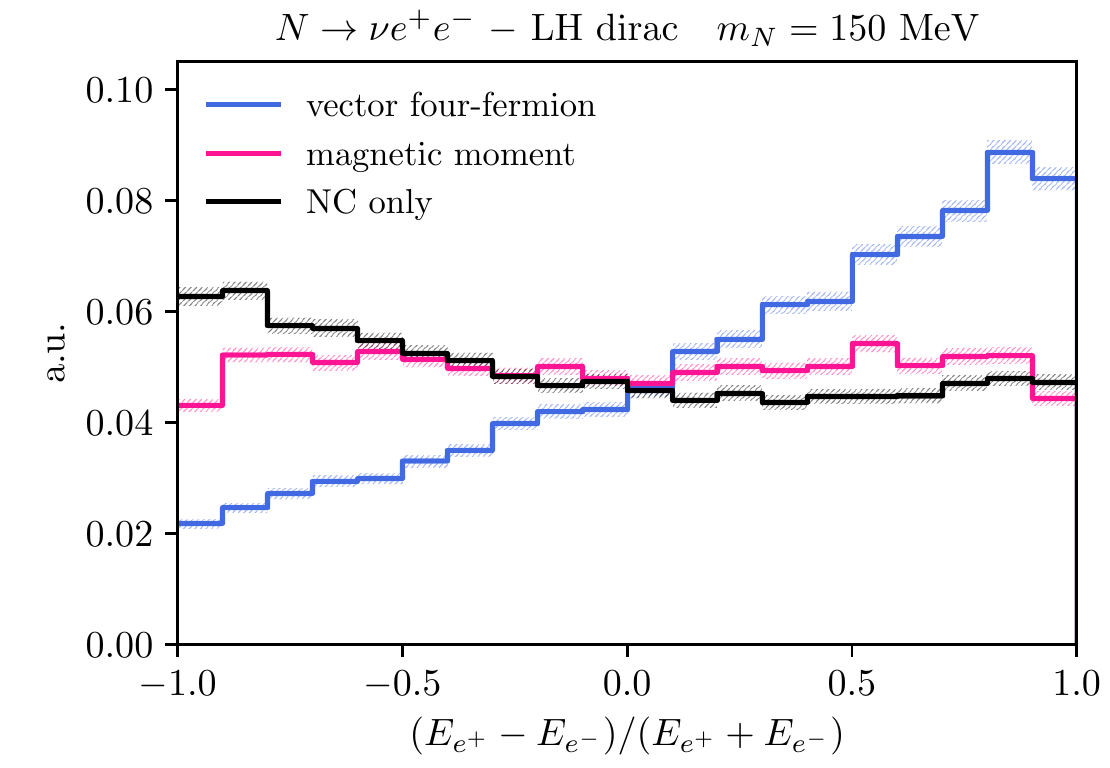}        \caption{Decay kinematics in different models. In addition to the main variables also shown in \cref{fig:distribution_masses}, we show the invariant mass, the angle of the sum of the $e^+e^-$ momenta with respect to the neutrino beam, and the energy asymmetry of the pairs.\label{fig:distributions_model}}
\end{figure*}

\subsubsection{Weak decays}
The weak decays of HNLs have been computed several times in the literature. Here we show the differential rate explicitly to clarify any differences between Dirac and Majorana, and decays occurring purely via CC, NC, or both. Neglecting the final lepton masses, we find
\begin{widetext}
\begin{align}\label{eq:decay_rates}
     \frac{\dd \Gamma^{\rm Dir}}{\dd E_+ \dd E_-} &= |U_{\alpha N}|^2 \frac{G_F^2 m_N}{2 \pi^3} \left[g_R^2 E_- (m_N - 2 E_-) + (1-g_L)^2 E_+(m_N-2E_+)\right],
\\
   \frac{\dd \Gamma^{\rm Maj}}{\dd E_+ \dd E_-} &= |U_{\alpha N}|^2 \frac{G_F^2 M_N}{2 \pi^3}  \left[(1+g_L)^2+g_R^2\right]\big[ m_N (E_+ + E_-) - 2(E_+^2 + E_-^2)\big],
\end{align}
\end{widetext}
where $g_L=\sin^2(\theta_W) - 1/2$, $g_R = \sin^2{\theta_W}$, and $E_{+}$ and $E_-$ are the positive and negative charged lepton energies in the center-of-mass frame, respectively. 
The expression holds for both left-handed and right-handed polarized HNLs. Note that the above rates in a purely CC decay mode can be recovered with $g_R,g_L\to0$, and in a purely NC decay mode with $(1+g_L)\to g_L$. The expression quoted in~\cite{Bernardi:1985ny} is in agreement with our calculation in the CC-only limit.

Note that in the Dirac case, the shape of the differential rate is sensitive to the interference between NC and CC. This effect is mostly irrelevant for variables of experimental interest. In \cref{fig:distribution_vertices}, we show a few kinematical variables for HNLs decaying inside ND280. Experimentally relevant variables are insensitive to the diagram responsible for the decay. The most visible difference is in the energy asymmetry (not shown).

\subsubsection{Transition Magnetic Moment}
For a large enough transition magnetic moment, HNLs will decay primarily via the two-body decay $N\to\nu\gamma$. For Dirac HNLs, the decay rate is
\begin{equation}
    \dd \Gamma_{N\to\nu\gamma}^{\rm Dir} = \frac{|\mu_{\rm tr}|^2 m_N^3}{16\pi},
\end{equation}
recalling that $\mu_{\rm tr} = 2d$, where $d$ is the dipole parameter often used in part of the literature~\cite{Magill:2018jla}.
Another possibility is the rate into a virtual photon. In that case, the dominant contribution comes from small photon virtualities, making $N\to \nu \gamma^* \to \nu e^+ e^-$ the dominant decay after the two-body one. Due to the small electron mass the branching ratio is enhanced by large logs and is larger than a naive factor of $\alpha/4\pi \times \mathcal{B}(N\to \nu \gamma)$. In fact, the differential rate will peak at the smallest lepton energies, as follows
\begin{equation}
    \frac{\dd\Gamma_{N\to \nu \gamma^* \to \nu \ell^+ \ell^-}^{\rm Dir}}{\dd E_+ \dd E_- } = \frac{\alpha |\mu_{\rm tr}|^2}{8\pi^2} \frac{m_N (E_++E_-) - 4E_+E_-}{(E_++E_-) - m_N/2},
\end{equation}
where we have taken the massless limit for the amplitude. In the massive case, the rate is regulated by the lepton mass to give
\begin{align}\label{eq:decay_rate_nuee_dipole}
        \Gamma_{N\to \nu \gamma^* \to \nu \ell^+ \ell^-}^{\rm Dir} &= \frac{\alpha |\mu_{\rm tr}|^2}{48\pi^2} m_N^3  L\left(\frac{m_\ell}{m_N}\right),
\end{align}
with
\begin{align}
    L(r) &= \left(2 - \frac{r^6}{8}\right) \,{\rm sech^{-1}}(r) - \frac{24 - 10r^2  + r^4}{8}\sqrt{1-4 r^2}
    \\\nonumber 
    &\simeq 2 \log\left(\frac{1}{r}\right) - 3  + \mathcal{O}(r^2),
\end{align}
where we show the leading-log approximation for small $r$ in the second line. As an example, for a HNL with $m_N = 100$~MeV and negligible mixing with active neutrinos, we get $\mathcal{B}(N\to\nu e^+e^-) = 0.68\%$. Above the dimuon threshold, we find $\mathcal{B}(N\to\nu e^+e^-) = 0.75\%$ and $\mathcal{B}(N\to\nu \mu^+\mu^-) = 3.8\times 10^{-3}$ for $m_N = 300$~MeV. At large masses, vector meson dominance dictates that $N$ will decay to final $\rho$, $\omega$ and $\phi$ mesons, which produce primarily pions and kaons accompanied by a final state neutrino.

The rate is peaked at the lowest dilepton invariant masses, implying that the signal will be very similar to real photons that convert in the detector material. The dependence of the rate on an artificial cut on $m_{\ell \ell}^2 = (p_{+} + p_-)^2$ can be understood analytically. At small values of $r_{ee} = m_{ee}^{\rm min}/m_N$, the rate will decrease as $\log(1/r_{ee})$, while for large $r_ee$, we expand in $a = 1 - r_{ee}^2$ to find a rate as in \cref{eq:decay_rate_nuee_dipole} with the replacement
\begin{align}
        L(r) \to a^3\sqrt{1-4r^2}\left(\frac{1}{2} + r^2 \right).
\end{align}

\subsubsection{Four-fermion vector interactions}

The decay via dim-6 operators will be analogous to the weak-decays case. The decay rate into leptons via the vectorial operator is given by
\begin{equation}
    \Gamma^{\rm Dir}_{N\to\nu (Z^\prime)^* \to \nu e^+e^-} = \frac{G_X^2 M_N^5}{192 \pi^3}.
\end{equation}
The differential rate in this case differs from the weak-decays only due to the lack of axial-vector couplings,
\begin{equation}
    \frac{\dd\Gamma_{N\to\nu \ell^+ \ell^-}^{\rm Dir}}{\dd E_+ \dd E_- } = 
    \frac{G_X^2 m_N}{32 \pi^3} \left(m_N (E_++E_-) - 2(E_+^2+E_-^2)\right).
\end{equation}
It is easy to see that this is the limit of \cref{eq:decay_rates} where CC is absent and $g_L=g_R$.

\section{HNL production from pion and muon decays}

\begin{figure*}[t]
    \centering
    \includegraphics[width=0.49\textwidth]{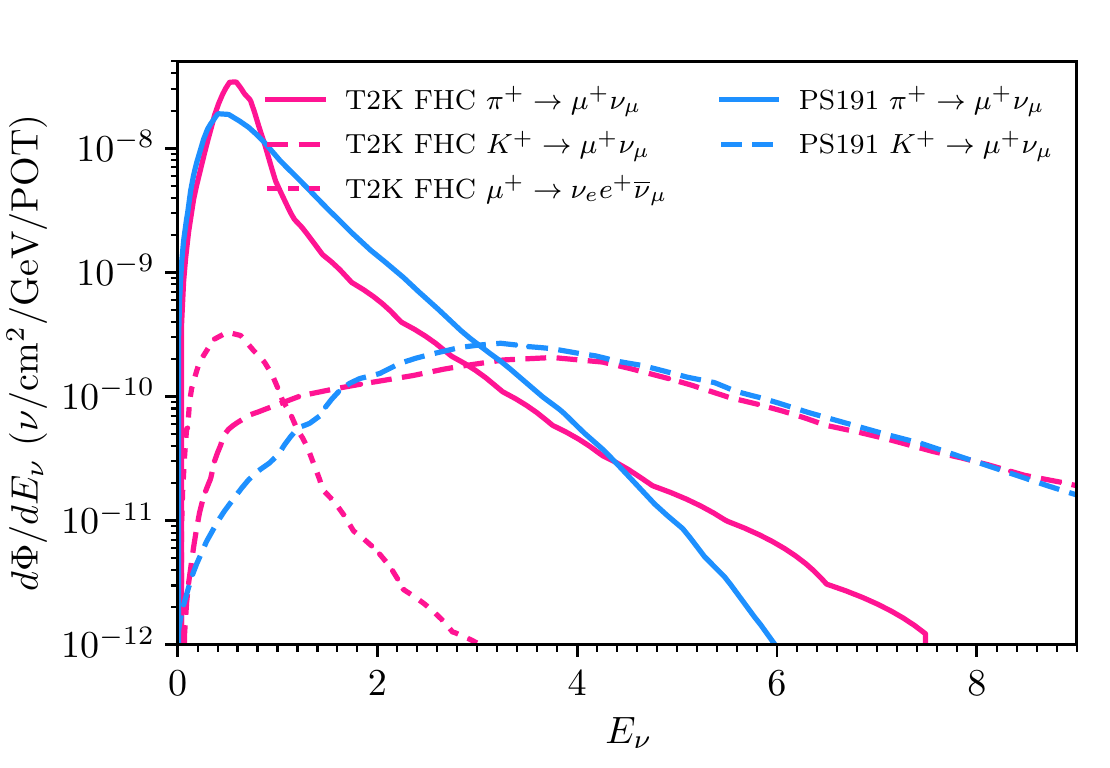}
    \includegraphics[width=0.49\textwidth]{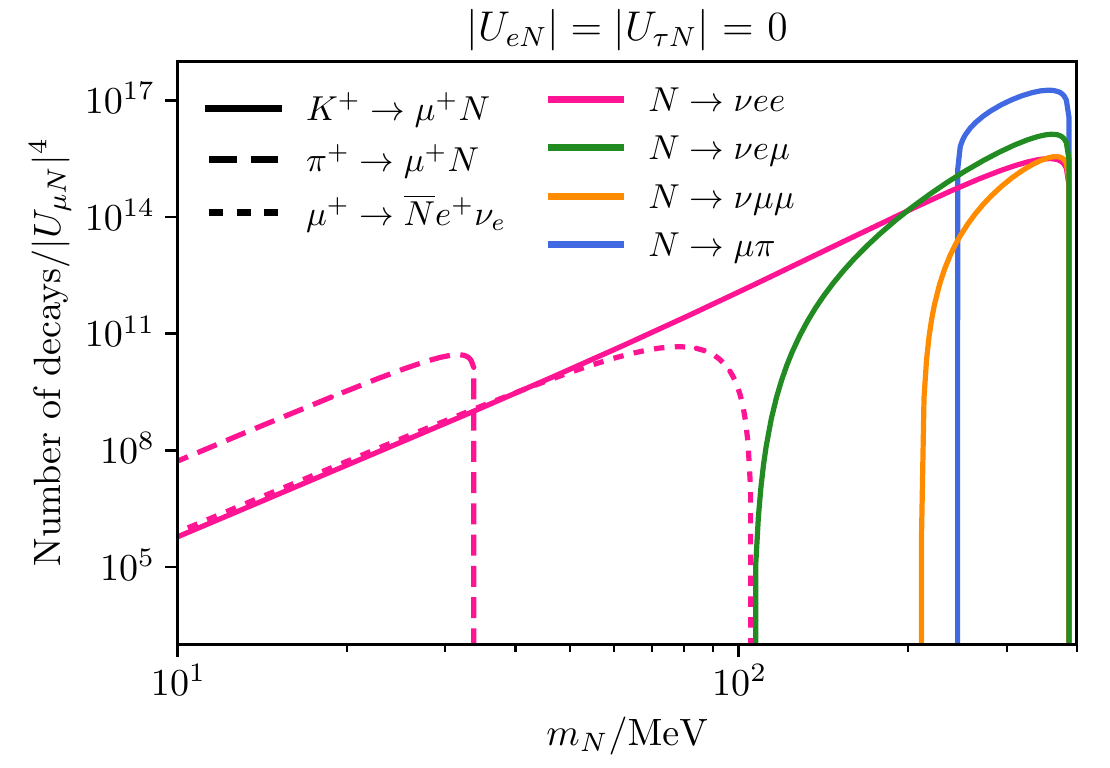}
    \caption{
    On the left panel we show the $\nu_\mu$ and $\overline{\nu}_\mu$ flux separated by parent particle at T2K in neutrino mode and PS-191 as a function of the neutrino energy. On the right, we show the number of decays in flight in the T2K ND280 detector for $18.63\times10^{20}$ as a function of the HNL mass. The number of decays shown assumes the long-lifetime limit and is normalized by $|U_{\mu 4}|^4$. We include an estimate of number of decays from HNLs produced in pion and muon decays for comparison.
    }
    \label{fig:hnl_flux_comparison}
\end{figure*}

We show the neutrino fluxes used in our main analysis on the left panel of \reffig{fig:hnl_flux_comparison}. The neutrino fluxes separated by parent particle are obtained from \refref{T2K:2012bge} for T2K, and \refref{Bernardi:1985} for PS191. We have checked that our fluxes agree reasonably well with those provided in Refs.~\cite{Chauveau:1985iz} and \cite{Bernardi:1986hs}.

Note that T2K did not include HNL production from pion and muon decays at the target. For this reason, we leave these decay channels out of our analysis since the final signal efficiencies can vary significantly between these channels due to the energy distribution and acceptance.
Nevertheless, we perform a naive comparison between the event rate in these channels and the one from kaon production, before efficiencies. Our results are shown on the right panel of \reffig{fig:hnl_flux_comparison}. 
To compute the HNL flux from muon decays, we take the approximation that the HNL flux is given by $\Phi_{\mu \to \overline{N} e\nu_e} \sim \Phi_{\mu\to\overline{\nu_\mu} e\nu_e} \times |U_{\mu 4}|^2 \times (\Gamma_{\mu\to \overline{N} e\nu_e}/\Gamma_{\mu \to \overline{\nu_\mu} e\nu_e})$, which is expected to be less accurate than in the case of the two-body meson decays used in the main text. 

\section{Existing limits from PS191}\label{app:ps191}

We revisit the constraints set by the PS191 experiment on the in-flight decays of heavy neutral leptons (HNL). We are primarily interested in the search for $K\to \mu (N \to \nu e^+e^-)$. As discussed in the main text, PS191 did not include neutral-current (NC) decays of HNLs, so this channel was used to constrain only the product $|U_{e N}U_{\mu N}|$. Including neutral currents, however, this limit can be naively translated to a limit on $|U_{\mu N}|^2$ as follows
\begin{equation}
    |U_{\mu N}|^2_{\rm new-limit} = \left(\frac{1 - 4s_{\rm W}^2 + 8s_{\rm W}^4}{4}\right)^{1/2}\times |U_{e N}U_{\mu N}|_{\rm PS191},
\end{equation}
where $s_{\rm W} = \sin{\theta_{\rm W}}$ is the sine of the weak mixing angle. 
The selection efficiency for charged-current (CC) and NC decays is expected to be similar. 
The bound obtained with this rescaling procedure and the one obtained with our simulations differ by a factor of $\approx 6$, which corresponds to a discrepancy by a factor of 36 in the event rate. 
We have checked that the naive estimates of the flux-averaged HNL decay rates at PS191 and T2K are of the same order, corroborating our conclusions.

\renewcommand{\arraystretch}{1.2}
\begin{table*}[t]
    \centering
    \begin{tabular}{|c|c|c|c|}
    \hline\hline
        & T2K & PS-191 & MicroBooNE (NuMI KDAR)\\
    \hline\hline
    $\langle E_{\nu^{\pi{\rm-decay}}} \rangle$     & $0.9$~GeV  &  $1$~GeV & ---  \\
    $\langle E_{\nu^{K\,{\rm-decay}}} \rangle$     & $4$~GeV  &  $4$~GeV & $234$~MeV  \\
    $\nu^{\pi{\rm -decay}}/$cm$^2$/POT   & $1.8\times10^{-8}$  & $1.7\times10^{-8}$ &  --- \\
    $\nu^{K{\rm -decay}}/$cm$^2$/POT   & $9.1\times10^{-10}$  & $1.0\times10^{-9}$ & $6.6\times10^{-11}$ \\
    POT    & $(12.34 + 6.29) \times 10^{20}$ &  $0.89\times10^{19}$ & $1.93\times 10^{20}$  \\
\hline \hline
    area & $1.7$~m$\times 1.96$~m & $3$~m$\times6$~m & $10.36$~m$\times 2.56$~m \\
    length    & $1.68$~m  & $12$~m & $3$~m \\
    baseline & $280$~m & $128$~m & $102$~m \\
    signal efficiency    & $6\% - 12\%$ & $\lesssim 30\%$ & $\sim 14\%$\\
\hline
\hline
    target & beryllium & carbon  & --- \\
    target length & $80$~cm & $91.4$~cm  & --- \\
    baseline & $280$~m & $128$~m     & --- \\
    decay tunnel & 96~m & 49.1~m & --- \\
    proton energy    & $30$~GeV  &  $19.2$~GeV & --- \\
    off-axis angle   & $2.042^\circ$  &  $2.29^\circ$     & --- \\
    \hline\hline
    \end{tabular}
    \caption{Comparison between T2K and PS191 experimental design. The top rows show numbers that enter directly in the overall normalization of the event rate. The quantity $\langle E_{K \to \nu} \rangle$ is defined as the average energy of neutrinos produced in kaon decays and $\nu/\text{cm}^2/\text{POT}$ is defined as the total neutrino flux integrated over all energies.}
    \label{tab:ps191_t2k_info}
\end{table*}
\renewcommand{\arraystretch}{1.0}

A direct comparison between the exposure of PS-191, which ran for a single month, and T2K, which ran for almost 7 years, corroborates our conclusions that PS-191 cannot significantly outperform T2K. A naive estimate for the total number of HNLs crossing the detector in the long-lifetime and ultra-relativistic limit, up to experiment-independent factors, is
\begin{equation}
    {\rm norm}  \simeq  \frac{n_{\rm POT}\times \Phi_{N} \times {\rm Area} \times {\rm Length} \times \langle \epsilon_{\rm sig}\rangle}{\langle E_{N}\rangle},
\end{equation}
where $\langle \dots \rangle$ denotes an average over the HNL spectrum, and $\epsilon$ the signal efficiency, and $\Phi_N$ the total flux of HNLs. A simple ratio between the two experiments using the information in \cref{tab:ps191_t2k_info} is
\begin{equation}
    \frac{({\rm norm})_{\rm PS-191}}{({\rm norm})_{\rm T2K}} \simeq 0.5.
\end{equation}
The numbers used for PS191 are obtained from Refs.~\cite{Chauveau:1985iz,Vannucci:1985vs,Bernardi:1985ny,Bernardi:1987ek}. We also note that T2K explores several analysis channels, including the efficiency to select $e^+e^-$ final states in other selection channels such as $\mu \pi$ and $e\pi$. This combination provides additional statistical power to the T2K analysis.

The most uncertain ingredient in our calculation of the normalization is the PS191 efficiency. From~\cite{Chauveau:1985iz}, we know that the detection efficiency is at most $70\%$. In addition, we are given the geometrical acceptance for $\pi$ final states, $\sim 40\%$, which in the low-density detector cannot be much different from the efficiency to detect electron final states. 
Therefore, $30\%$ is the largest efficiency we allow for PS-191 to have, assuming it remains constant for all HNL masses. 
The real efficiency is likely to be smaller, however, especially at low HNL masses where the $e^+e^-$ final state is more collimated. 
In any case, even for $100\%$ efficiencies, we do not find reasonable agreement with the published bounds.

We have performed this check for the $\mu \pi$ and $\nu e \mu$ channels, finding similar discrepancies. For low HNL masses, we observe that the ratio between our bounds and the published PS191 ones goes from $\sim 6$ to lower values. This is most likely due to our assumption that the efficiencies remain constant.

\subsection{Comparison between T2K and MicroBooNE}

\begin{figure}
    \includegraphics[width=0.49\textwidth]{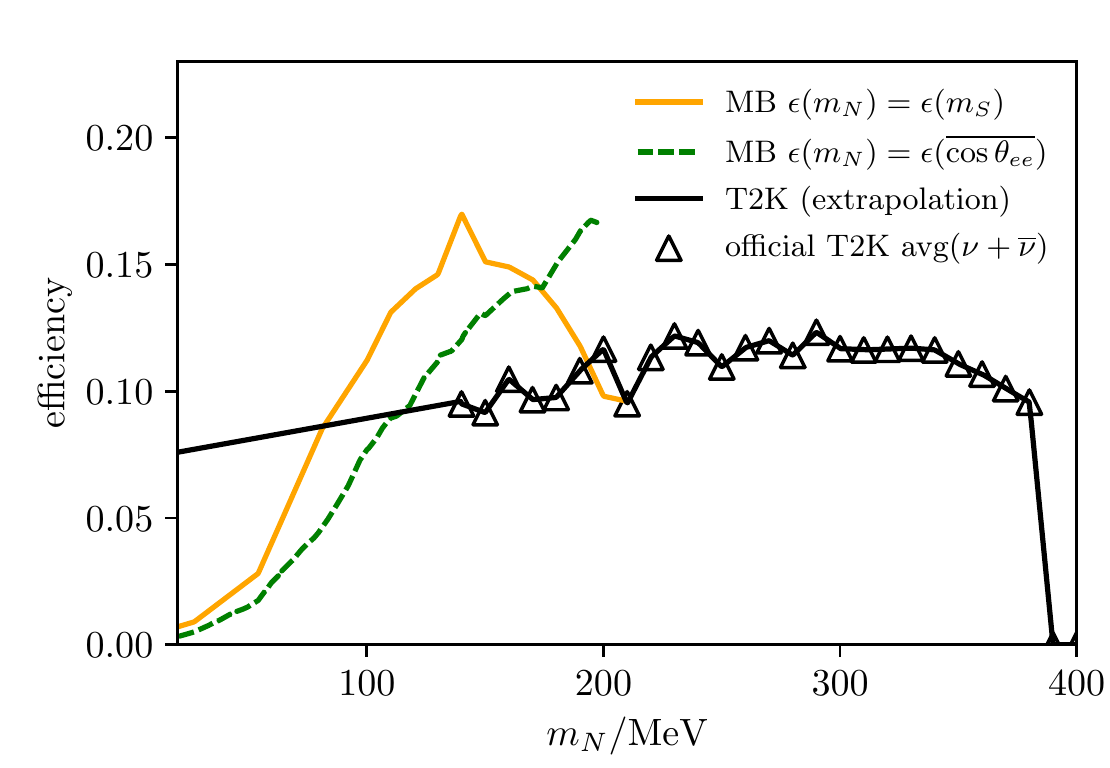}
    \caption{Comparison between the MicroBooNE and T2K efficiencies used in this work. Black triangles indicate the official efficiencies quoted in \refref{T2KND280TPC:2010nnd}, averaged according to the POT in neutrino and antineutrino mode. The green curve has been obtained from the analysis of Ref.~\cite{Kelly:2021xbv}. \label{fig:efficiencies}}
\end{figure}

\Cref{tab:ps191_t2k_info} also helps compare our T2K limits and those set by the authors of~\refref{Kelly:2021xbv} using the MicroBooNE analysis in~\refref{MicroBooNE:2021usw}. In the long-lived HNL limit, the ratio between the naive normalization factors in T2K and MicroBooNE is approximately
\begin{equation}
    \frac{({\rm norm})_{\rm MicroBooNE}}{({\rm norm})_{\rm T2K}} \simeq  1.8 \times \frac{\epsilon_{\rm MicroBooNE}}{\epsilon_{\rm T2K}},
\end{equation}
were the difference in detector size is compensated by the larger neutrino flux at T2K. As can been seen in \cref{fig:efficiencies}, the MicroBooNE efficiencies fall rapidly at low HNL masses. While the T2K efficiencies we use at low masses are obtained from an extrapolation, it is clear that their dependence on $m_N$ is much milder, as is also the case in other reconstruction channels, such as in $\mu^+\mu^-$ or $\mu^\pm e^\mp$. This is due to the low-density material, which prevents showering of the electrons, and the magnetic field, which splits the $e^+e^-$ pairs with measurable angles. A great example of the capability of ND280 to reconstruct the highly-collinear dilepton pairs is the single photon selection in the $\nu_e$CC measurements~\cite{T2K:2014djs,T2K:2020lrr} and in the single photon search~\cite{T2K:2019odo}. In the former, the selection efficiency for photons converted in the FGD was found to be $12\%$. In MicroBooNE, on the other hand, low-mass HNLs that decay into overlapping $e^+e^-$ pairs are much less likely to be reconstructed as two separate objects. More detailed analyses by the collaborations can refine our estimates for the efficiencies at T2K and improve on the numbers shown in \cref{tab:ps191_t2k_info}.

\subsection{Comparison in the light scalar case}

It is also possible to translate between bounds on the HNL mixing to bounds on the mixing of light scalars with the SM Higgs. In particular, due to the proximity between the muon and pion masses, the bounds on the scalar mixing can be approximately related to those on the HNL mixing under a muon-dominance assumption as follows:
\begin{equation}\label{eq:translate_scalar}
    \theta_{\rm bound}^4 = |U_{\mu N}|_{\rm bound}^4 \times \frac{\mathcal{B}(K^+ \to \mu^+ N)}{\mathcal{B}(K^+ \to \pi^+ \phi)} \frac{\hat{\Gamma}_{N\to \nu e^+e^-}}{\hat{\Gamma}_{\phi \to e^+e^-}},
\end{equation}
where $\hat{\Gamma}$ is the decay rate normalized by the relevant mixing angle. This neglects the contribution from $K_L$ as well as $K^-$ decays, as well as hadron regeneration, and so provides a conservative estimate for the bound on $\theta$. These effects, can contribute as much as a factor of 2 to the total rate, since 0.065 $K^+$, 0.032 $K^-$, and 0.044 $K_L$ are produced per proton on target according to \cite{Gorbunov:2021ccu}. For our PS-191 constraint, \refeq{eq:translate_scalar} gives $\theta^2 < 2.8 \times 10^{-7}$ for $m_{\phi} = 150$~MeV. Including a naive factor of 2 in the rate for the other kaon sources this translates to $\theta^2 < 2 \times 10^{-7}$. These values are not far from the constraints found in \cite{Gorbunov:2021ccu}. In fact, using the naively rescaled PS191 constraints on HNLs, we would have found $\theta < 4\times 10^{-8}$, which is much stronger than the quoted value in \cite{Gorbunov:2021ccu}.

\subsection{Reproducing the T2K limits}

\begin{figure*}[t]
    \centering
    \includegraphics[width=0.55\textwidth]{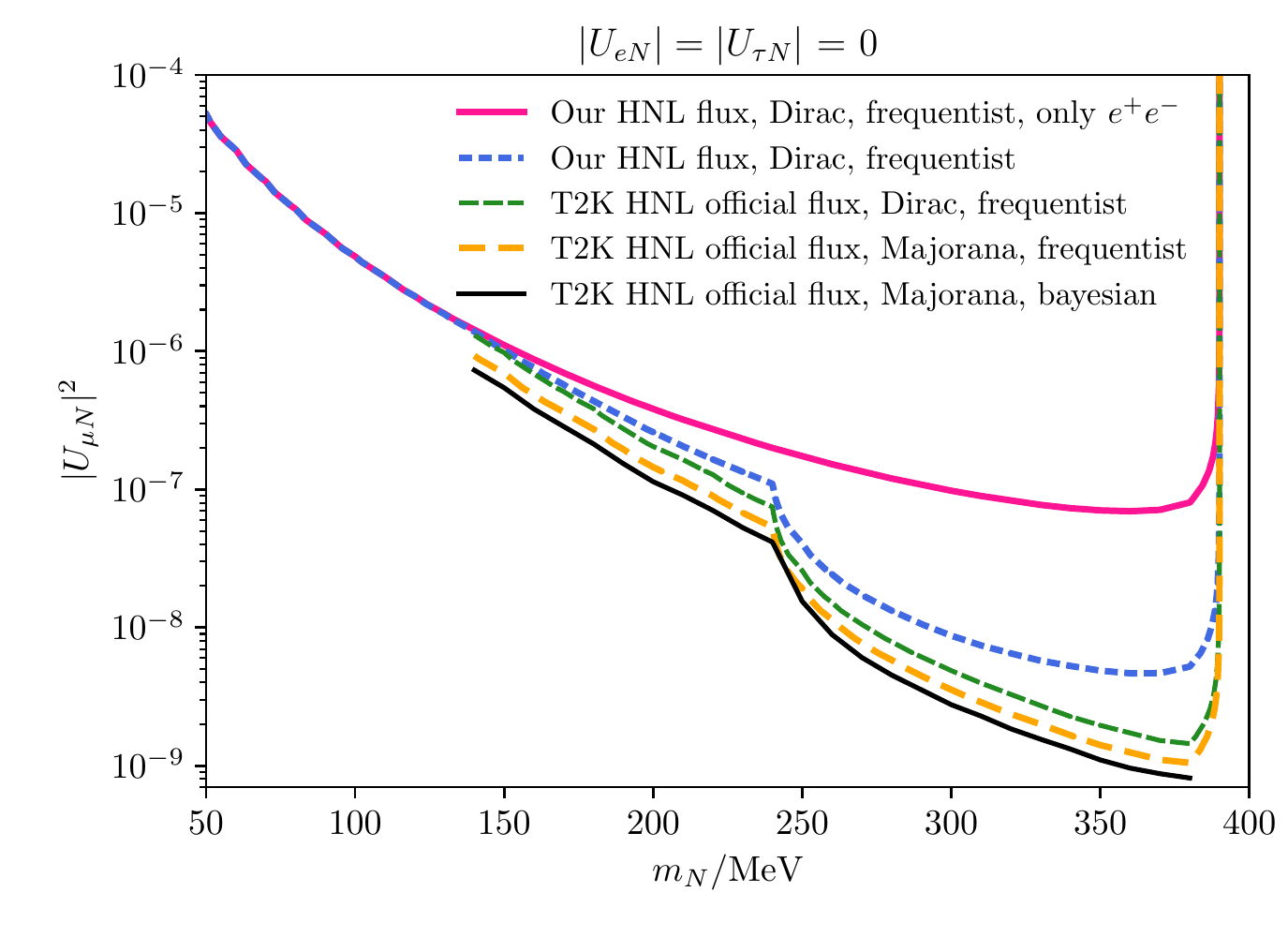}
    \caption{
    Several differences between our analysis and the T2K official result lead to a conservative limit.
    The T2K official result sets limit for Majorana neutrinos in a Bayesian fashion (black solid), which is stronger than using a frequentist limit (orange dashed), and is about a factor of $\sqrt{2}$ stronger than the limit for Dirac neutrinos (green dashed).
    Our flux estimation is conservative at large HNL mass, and matches the full simulation at lower masses (blue dashed). 
    Excluding all decay modes aside for $e^+e^-$ (pink line) leads to a conservative limit above the muon mass, and matches well with the equivalent T2K constraints below the pion mass.
    }
    \label{fig:t2k_limit_comparison}
\end{figure*}

We now discuss the consistency between our limits in~\cref{fig:minimal} and the official T2K result~\cite{Abe:2019kgx}.
A direct comparison between our limit and the figures in~\cite{Abe:2019kgx} would not be fair because of the different physical and methodological assumptions.
\Cref{fig:t2k_limit_comparison} illustrates the individual effect of each different ingredient and in what limit we reproduce the T2K official result correctly.

First, T2K performs a Bayesian analysis for Majorana neutrinos, while we are showing a frequentist analysis for Dirac neutrinos.
Moreover, T2K shows both marginalized limits, where the posterior is integrated over the parameters not shown in the current figure ($U_{eN}$ and $U_{\tau N}$ for our case), and profiled limits where the posterior density is shown after conditioning on the values of the parameters not shown equal to zero ($U_{eN} = U_{\tau N} = 0 $).
The second case is appropriate for our comparison, and we were able to plot it from the data release (black solid line).
From the data release, we also obtained the flux and the efficiency in the mass range considered for the analysis, from which we extracted a frequentist limit without considering systematics and backgrounds (orange dashed line), which play a minor role in this analysis.
The difference between a Dirac and a Majorana (green dashed curve) analysis is a factor of two in the HNL rate --- aside for a small difference in efficiency between charge conjugate channels for $e \mu$, $e \pi$, and $\pi \mu$ channels --- and becomes exactly a factor of two for the $e^+e^-$ channel.
We reproduce the T2K HNL flux using \cref{eqn:hnl_flux_scaling}, which largely underestimates the flux at large masses, but becomes accurate within 10\% at $m_N = \SI{150}\MeV$ (blue dashed line curve).
Eventually, considering only the $e^+e^-$ decay mode gives the most conservative limit as we are not adding contributions from $e\mu$, $e\pi$, $\pi \mu$, and $\mu \mu$.
However, below the muon mass, since $e^+e^-$ is the only kinematically allowed decay mode, the limit matches the case where all decay modes are considered, as shown by comparing the pink solid line with the blue dashed line.

\bibliographystyle{apsrev4-1}
\bibliography{lib}{}

\end{document}